\documentclass[a4paper,fleqn]{cas-sc}
 \ExplSyntaxOn \cs_gset:Npn \__first_footerline: { \group_begin: \small \sffamily \__short_authors: \group_end: } \ExplSyntaxOff

\usepackage[T1]{fontenc}
\usepackage[ansinew]{inputenc}
\usepackage{multirow} 
\usepackage[english]{babel}
\usepackage{graphicx}
\usepackage{amsmath,bm}
\usepackage{amssymb}
\usepackage{eso-pic}
\usepackage{fancyhdr}
\usepackage{pdfpages}
\usepackage{subfig}
\usepackage{changepage}
\usepackage{setspace}
\usepackage{multirow}
\usepackage{framed}
\usepackage{geometry}
\usepackage[justification=justified]{caption}
\usepackage[toc,page,titletoc]{appendix}
\usepackage{csquotes}
\usepackage{tikz}
\usetikzlibrary{shapes.geometric, arrows}
\usetikzlibrary{matrix}
\usetikzlibrary{fit}
\usetikzlibrary{backgrounds}
\usepackage{natbib}
\usepackage{floatrow}
\usepackage{times}
\usepackage{bm}
\usepackage{color}
\usepackage{makecell}
\usepackage{tabularx}
\usepackage{soul}
\usepackage[plain,noend]{algorithm2e}

\def\tsc#1{\csdef{#1}{\textsc{\lowercase{#1}}\xspace}}
\tsc{WGM}
\tsc{QE}

\newcommand{\iid}{\stackrel{\rm iid}{\sim}}

\appendixtitleon 
\newfloatcommand{capbtabbox}{table}[][\FBwidth]

\graphicspath{{figures/}}

\begin{document}
\title[mode=title]{Bayesian modeling and clustering for spatio-temporal areal data:\\ An application to Italian unemployment}  
\shorttitle{Bayesian modeling and clustering for spatio-temporal areal data: An application to Italian unemployment}

\author[1]{Alexander Mozdzen}[orcid=0000-0002-4361-3881]
\cormark[1]

\ead{alexander.mozdzen@aau.at}
\author[2]{Andrea Cremaschi}[orcid =0000-0003-1179-0768]
\ead{cremaschia@sics.a-star.edu.sg}
\author[1]{Annalisa Cadonna}[orcid=0000-0003-0360-7628]
\ead{annalisa.cadonna@aau.at}
\author[3]{Alessandra Guglielmi}[orcid=0000-0001-7005-7588]
\ead{alessandra.guglielmi@polimi.it}
\author[1]{Gregor Kastner}[orcid=0000-0002-8237-8271]
\ead{gregor.kastner@aau.at}

\shortauthors{A. Mozdzen, A. Cremaschi, A. Cadonna, A. Guglielmi and G. Kastner}

\cortext[1]{Corresponding author at: Department of Statistics, University of Klagenfurt, Universit\"atsstra\ss{}e 65-67, 9020 Klagenfurt am W\"orthersee, Austria.}
\nonumnote{Alexander Mozdzen, Annalisa Cadonna and Gregor Kastner acknowledge funding from the Austrian Science Fund (FWF) for the project
``High-dimensional statistical learning: New methods to advance economic and sustainability
policy'' (ZK 35), jointly carried out by the University of Klagenfurt, Paris Lodron University Salzburg, TU Wien, and the Austrian
Institute of Economic Research (WIFO).}

\address[1]{Department of Statistics, University of Klagenfurt, Klagenfurt, Austria}
\address[2]{Singapore Institute for Clinical Sciences (SICS), A*STAR, Singapore}
\address[3]{Department of Mathematics, Politecnico di Milano, Milano, Italy}

\begin{abstract}\noindent
Spatio-temporal areal data can be seen as a collection of time series which are spatially correlated according to a specific neighboring structure. Incorporating the temporal and spatial dimension into a statistical model poses challenges regarding the underlying theoretical framework as well as the implementation of efficient computational methods. We propose to include spatio-temporal random effects using a conditional autoregressive prior, where the temporal correlation is modeled through an autoregressive mean decomposition and the spatial correlation by the precision matrix inheriting the neighboring structure. Their joint distribution constitutes a Gaussian Markov random field, whose sparse precision matrix enables the usage of efficient sampling algorithms. We cluster the areal units using a nonparametric prior, thereby learning latent partitions of the areal units. The performance of the model is assessed via an application to study regional unemployment patterns in Italy. When compared to other spatial and spatio-temporal competitors, the proposed model shows more precise estimates and the additional information obtained from the clustering allows for an extended economic interpretation of the unemployment rates of the Italian provinces.
\end{abstract}


\maketitle

\section{Introduction}
\label{sec:intro}

Data representing spatial features collected at different time points have witnessed an increase in availability in recent years. The underlying areal units can be defined by geographical boundaries (e.g., regions, countries, municipalities) or by a tessellation of the territory of interest. This type of spatio-temporal data is commonly encountered in fields such as Social Sciences, Telecommunications, Economics, Epidemiology, and Image Analysis. A current example is the count of COVID-19 registered cases, often communicated on a daily basis per country and region.
We can therefore think of spatio-temporal areal data as a collection of time series, whose spatial correlation depends on the geographical location of the underlying areal units.

In this work, we focus on the analysis of data regarding the evolution of unemployment rates in Italy. Italy has one of the highest unemployment rates in the European Union and exhibits well-known economic disparities between the northern and the southern regions \citep[see][]{it_unemp_brunello}. The global economic and the following eurozone crisis had a destructive impact on the Italian economy, leading to a sharp rise of unemployment across the entire country \citep[see][]{italy_global_crisis,italy_eurozone_crisis}. 
The geographical dichotomy as well as the upward trend of the unemployment rate
motivates the need for a flexible modeling approach, explicitly incorporating both spatial and temporal dimensions.

In addition to modeling the data in space and time, we are interested in uncovering economic patterns across provinces, allowing for the identification of groups of areal units sharing similar features. To this end, we adopt a model-based clustering approach exploiting tools from the Bayesian nonparametric (BNP) literature. More specifically, we make use of the popular Dirichlet Process (DP) prior \citep{ferguson1973bayesian}, which avoids the need of fixing the number of clusters in the model. See \cite{quintana2003bayesian} for the relationship between model-based clustering and the DP prior.

The body of literature on Bayesian spatio-temporal models for areal data in regional science and quantitative spatial analysis is large. Pars pro toto, we mention the textbooks by \cite{Banerjee2014}, \cite{haining2020regression}, and \cite{sahu2022bayesian}.
A widely used class of models for the study of areal data is that of conditional autoregressive (CAR) models, introduced in \cite{besag1974spatial, Besag1975}. In a CAR model, the spatial dependence among the observations is captured by a correlation matrix containing information on the neighboring structure of the data.
This approach can be used to define a prior distribution for random effects in a Bayesian hierarchical model. Bayesian hierarchical models are particularly useful in modeling related observations, such as in the case of areal time-varying data, allowing to capture the underlying dependence structures and providing full inference and uncertainty quantification. Different CAR prior specifications can be found in the literature, such as the intrinsic-CAR (ICAR) and Besag-York-Molli\'e priors \citep[both in][]{Besag1991}. Here, we follow \cite{Leroux2000}, who specify a set of spatially correlated random effects within a generalized regression setting. CAR models can be thought of as the conditional specification of a (Gaussian) Markov random field (GMRF). The different CAR models proposed in the literature correspond to a specific choice of the precision matrix of the corresponding GMRF \citep[see][for more details]{Rue2005}. Thanks to the Markov property, the precision matrix of a GMRF is usually sparse, enabling efficient computations through algorithms for sparse matrices based on results from linear algebra. An algorithm for block updating for Markov random fields models can be found in \cite{knorr-held2002}.

In a Bayesian framework, approaches for dealing with both spatial and temporal dependence in areal data can be found, among others, in \cite{UGARTE2012}, \cite{RUSHWORTH201429}, \cite{lawson2013bayesian}, and \cite{Napier_2018}.
\cite{beraha2021spatially} consider the problem of spatially dependent areal data and propose modeling the data collected for each areal unit through a finite mixture of Gaussian distributions. The spatial dependence is introduced via a joint distribution for a collection of vectors on the simplex, which is a logistic transformation of Gaussian multivariate CAR models.
\cite{nicoletta2022bayesian} propose a Bayesian model for spatio-temporal data based on a generalized linear mixed effects model. In the spirit of the current paper, \cite{fischer2015ab} calculate  the posterior predictive probabilities for the assignment of areal units to clusters using a spatio-temporal stochastic panel relationship of economic variables such as income and human capital.
 
Of particular interest is the model by \cite{lee2016}, introducing temporal dependence among areal units through an autoregressive structure on the means of the GMRF over time. The model in \cite{lee2016} allows for clustering by modeling the spatio-temporal parameters through a parametric mixture. The model proposed in this work can be seen as an extension of \cite{lee2016}. We introduce location-specific autoregressive parameters and use a nonparametric prior for model-based clustering of the areal-specific time series. In particular, we employ the popular Dirichlet process prior \citep{ferguson1973bayesian} to jointly model the effects of the time- and areal-specific covariates on the evolution of the unemployment rates and the autoregressive coefficients used to specify the distribution of the spatially correlated random effects. From a computational point of view, we exploit the fact that the proposed spatio-temporal model can be seen as a GMRF and combine the algorithms in \cite{knorr-held2002} and \cite{Mccausland2011} to propose an MCMC algorithm for efficient posterior simulation. 

A preliminary version of the model proposed in this work can be found in \cite{cadonna2019bayesian}, where the authors study the use of mobile phones in the municipality of Milan with the goal of investigating the local population density dynamics. However, while they use harmonic components which are constant across the areal units, we include time- and areal-specific predictors and cluster the areal units. Moreover, the spatial domain in \cite{cadonna2019bayesian} is a regular bivariate grid of the metropolitan area of Milan, while here the proximity matrix is given by neighboring Italian administrative provinces.
We point out that we introduce a Bayesian model for clustering the areal-specific time series of the unemployment rates with cluster estimates that do not vary with time. Time-varying clustering of the provinces would require a different modeling approach, see, for instance, \cite{page2021dependent}, and \cite{deiorio2019bayesian}. \cite{nieto2014bayesian}, instead, use subject-specific parameters from a BNP prior to cluster time series of share prices in the Mexican stock exchange.

The main contributions of this paper to the analysis of spatio-temporal areal data are the following: (a) perform model-based clustering of areal time series through the effects of time- and areal-specific predictors as well as CAR-modeled random effects; (b) account for spatio-temporal random effects using the CAR specification; (c) implement an efficient MCMC algorithm based on the GMRF interpretation of the CAR prior by exploiting the sparse structure of the spatial correlation matrix; (d) provide interpretable estimates of the evolution of unemployment rates in Italy and its connection to economic factors; (e) offer a thorough comparison of the proposed model with competitors, illustrating the advantages of the proposed approach. This work shows the merits of Bayesian modeling in terms of interpretability, outperforming existing complex nonlinear models such as the one in \cite{Minguez2020} and can help economists and policy-makers to better understand the evolution of important macroeconomic quantities.

The remainder of this paper is organized as follows. Section~\ref{sec:data} describes the motivating application and introduces a preliminary exploratory analysis of the data under study. In Section~\ref{sec:Model}, we present the spatio-temporal clustering model including random effects and a BNP prior, which induces clustering through a mixture of the spatio-temporal parameters. Section~\ref{sec:algorithm} displays a sketch of the MCMC algorithm used for posterior inference and Section~\ref{sec:emp_app} presents the results of the empirical application. Finally, Section~\ref{sec:conclusion} concludes the paper with a discussion.
Appendices~\ref{sec:st_randeff} and~\ref{sec:fullcond_bnp_remain} provide details on the derivation of the full conditional distributions and the MCMC algorithm. Appendix~\ref{sec:sens_loss} shows the results of an extensive sensitivity analysis assessing the effect of the method of loss calculation used to estimate the partition of the areal units. Appendix~\ref{sec:alt_model_spec} provides a comparison of the proposed model with alternative model specifications. Finally, Appendix~\ref{sec:sim_study} presents the results of a simulation study with synthetic data.

\section{Data and exploratory analysis}
\label{sec:data}
In this section, we illustrate the data motivating the development of the proposed approach. We base the choice of dataset on \cite{Minguez2020}, who approach the problem of modeling spatio-temporal data using a semiparametric model including P-splines. The authors provide a thorough economic background for the choice of data, which is briefly reviewed here. First, we describe the response variable used in the analysis, the unemployment rate in Italy, and how it varies over time and across provinces. Then, we introduce the explanatory variables and elaborate on their correlations and distinctions between the North and the South.

\subsection{Unemployment Data}
\begin{figure}[t]
	\centering
		\includegraphics[width=.5\textwidth]{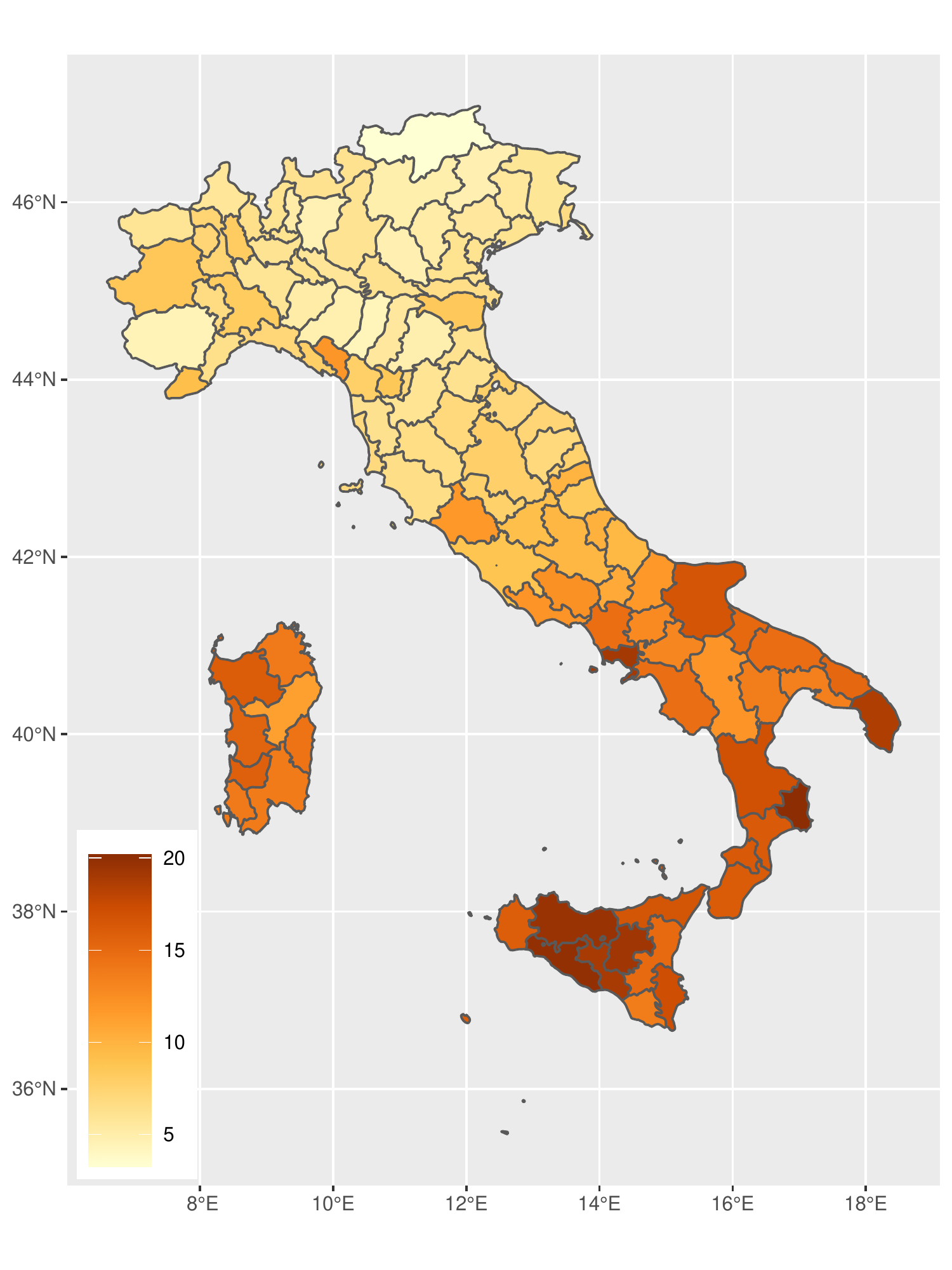}
		\caption{Average unemployment rates in Italian provinces from 2005 to 2017.}
		\label{fig:choropleth}
\end{figure}

\begin{figure}[t]
	\centering
		\includegraphics[scale=0.5]{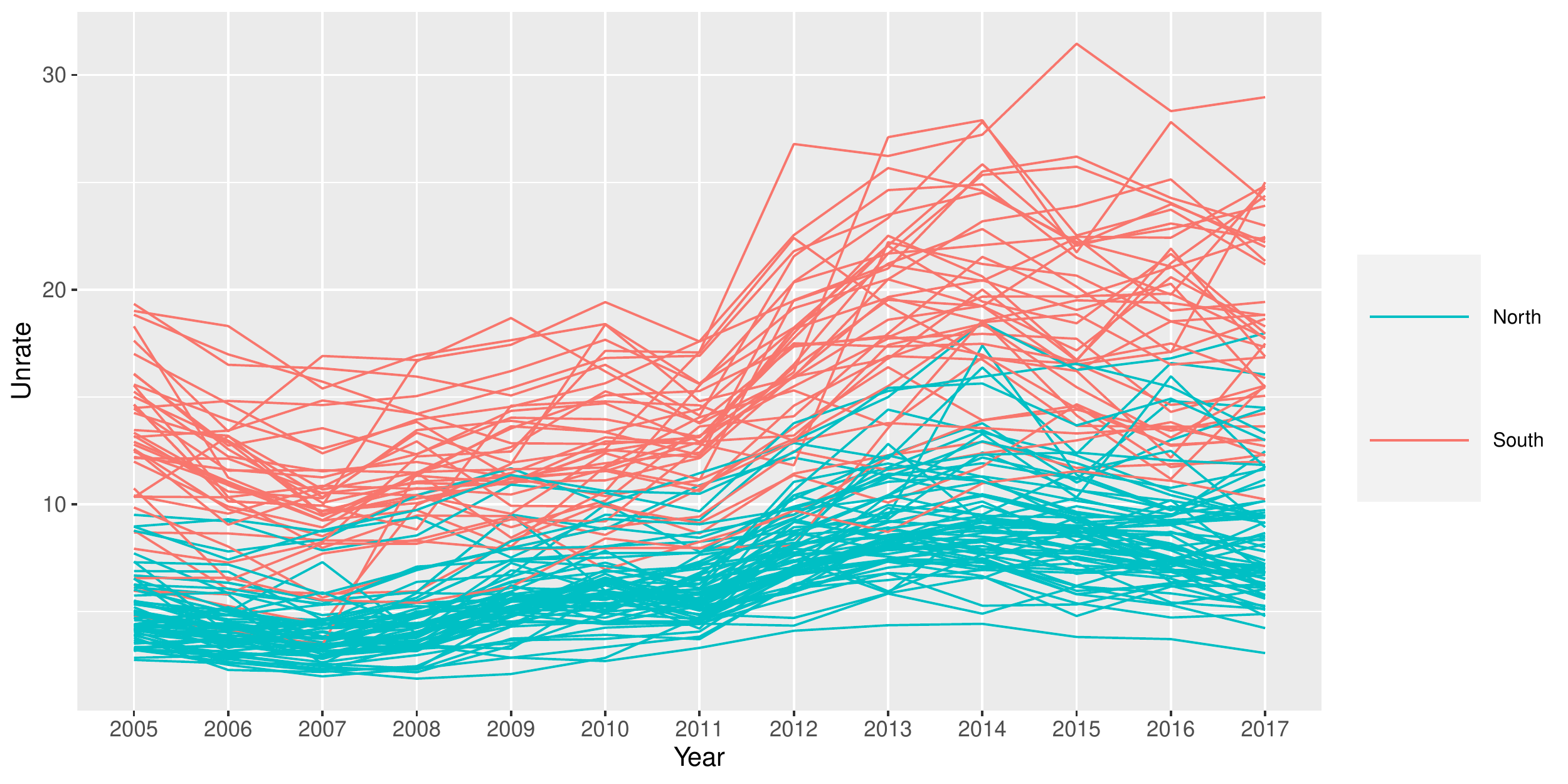}
		\caption{Unemployment rates over time from year 2005 to 2017.}
		\label{fig:NorthS_2}
\end{figure}

This study focuses on $I = 110$ Italian provinces and on $T = 13$ years from 2005 to 2017. Specifically, we use the third level of the ``nomenclature des unit\`es territoriales statistiques'' (NUTS) hierarchy. For each Italian province and each year, provincial unemployment rates are provided by the Italian National Institute of Statistics (ISTAT). More specifically, for province $i$ at year $t$, the unemployment rate is defined as $unrate_{i,t} = 100 \times {U_{i,t}}/{LF_{i,t}},$ where $U_{i,t}$ is the number of unemployed people and $LF_{i,t}$ is the labor force, including both the employed and unemployed. Figure~\ref{fig:choropleth} shows the average magnitude of the unemployment rate across Italian provinces. The economic differences between northern and southern provinces are clearly visible. This dichotomy has further increased over the years from 2012 to 2017, as can be observed in Figure~\ref{fig:NorthS_2}, displaying the evolution of unemployment rates from $2005$ to $2017$. Apart from distinctively higher unemployment rates, we additionally see a stronger upwards trend in recent years for the southern provinces. We draw the border between the North and the South according to the first level of the NUTS hierarchy above the regions Abruzzo, Molise, and Campania.

As exploratory measures of spatial autocorrelation we compute Moran's I \citep[MI,][]{moran} and Geary's C \citep[GC,][]{geary}. These
can be interpreted as the spatial couterpart of the lagged autocorrelation coefficient. MI ranges from $-1$ to $1$, indicating negative and positive spatial autocorrelation, respectively. Values of GC between 0 and 1 indicate positive, and values above 1 negative spatial autocorrelation. For the raw data, we obtain an average value over all years of $\text{MI} = 0.78$ and $\text{GC} = 0.21$, both values signifying strong positive spatial autocorrelation. Assuming differences between northern and southern provinces, we regress the unemployment rates on the latitude of the provinces and compute both measures on the residuals. The new results are $\text{MI} = 0.21$ and $\text{GC} = 0.76$. We conclude that when controlling for the known dichotomy, the statistics are significantly lower, but still indicate spatial autocorrelation in the data.

\subsection{Explanatory variables}
As mentioned before, the choice of explanatory variables follows \cite{Minguez2020}. Seven explanatory variables are selected for the econometric analysis, representing either so-called equilibrium or disequilibrium factors \citep{Rickman}. The variable representing the disequilibrium view is the employment growth rate ($empgrowth$), defined as the annual change of employment for each province. The remaining variables account for the equilibrium view. In order to capture the economic structure of the individual provinces, we include the share of people working in the economic sectors agriculture ($agri$), industry ($ind$), construction ($cons$), and services ($serv$). We distinguish between urban and rural areas by including the logarithm of the population density ($lpopdens$). The participation rate ($partrate$) is defined as the ratio between the total labor force and the working population and serves as a proxy for labor supply. Table \ref{tab:desc} shows summary statistics of the variables used in this work.
\begin{table}[t] 
		\begin{tabular}{@{\extracolsep{5pt}}lccccccc} 
			\\[-1.8ex]\hline 
			\hline \\[-1.8ex] 
			variable & \multicolumn{1}{c}{Mean} & \multicolumn{1}{c}{St. Dev.} & \multicolumn{1}{c}{Min} & \multicolumn{1}{c}{1st Qu.} & \multicolumn{1}{c}{3rd Qu.} & \multicolumn{1}{c}{Max} \\ 
			\hline \\[-1.8ex] 
			unrate & 9.942 & 5.484 & 1.873 & 5.774 & 12.927 & 31.456 \\ 
			\hline 
            agri & 5.229 & 4.017 & 0.037 & 2.227 & 7.416 & 24.782 \\ 
            ind & 20.252 & 8.902 & 4.486 & 12.908 & 26.531 & 47.86 \\ 
            cons & 8.142 & 2.178 & 3.242 & 6.481 & 9.525 & 16.579 \\ 
            serv & 66.366 & 7.674 & 45.155 & 61.351 & 71.624 & 88.15 \\ 
            partrate & 63.42 & 8.124 & 40.615 & 56.584 & 69.746 & 76.058 \\ 
            empgrowth & -0.006 & 4.222 & -45.892 & -1.689 & 1.924 & 60.524 \\ 
            lpopdens & -1.762 & 0.809 & -3.482 & -2.271 & -1.304 & 0.961 \\ 
			\hline \\[-1.8ex] 
		\end{tabular}
	\caption{Summary statistics of the response and predictors.} 
	\label{tab:desc}
\end{table}

\begin{figure}[t]
	\begin{center}
		\includegraphics[height=7cm,width=14cm]{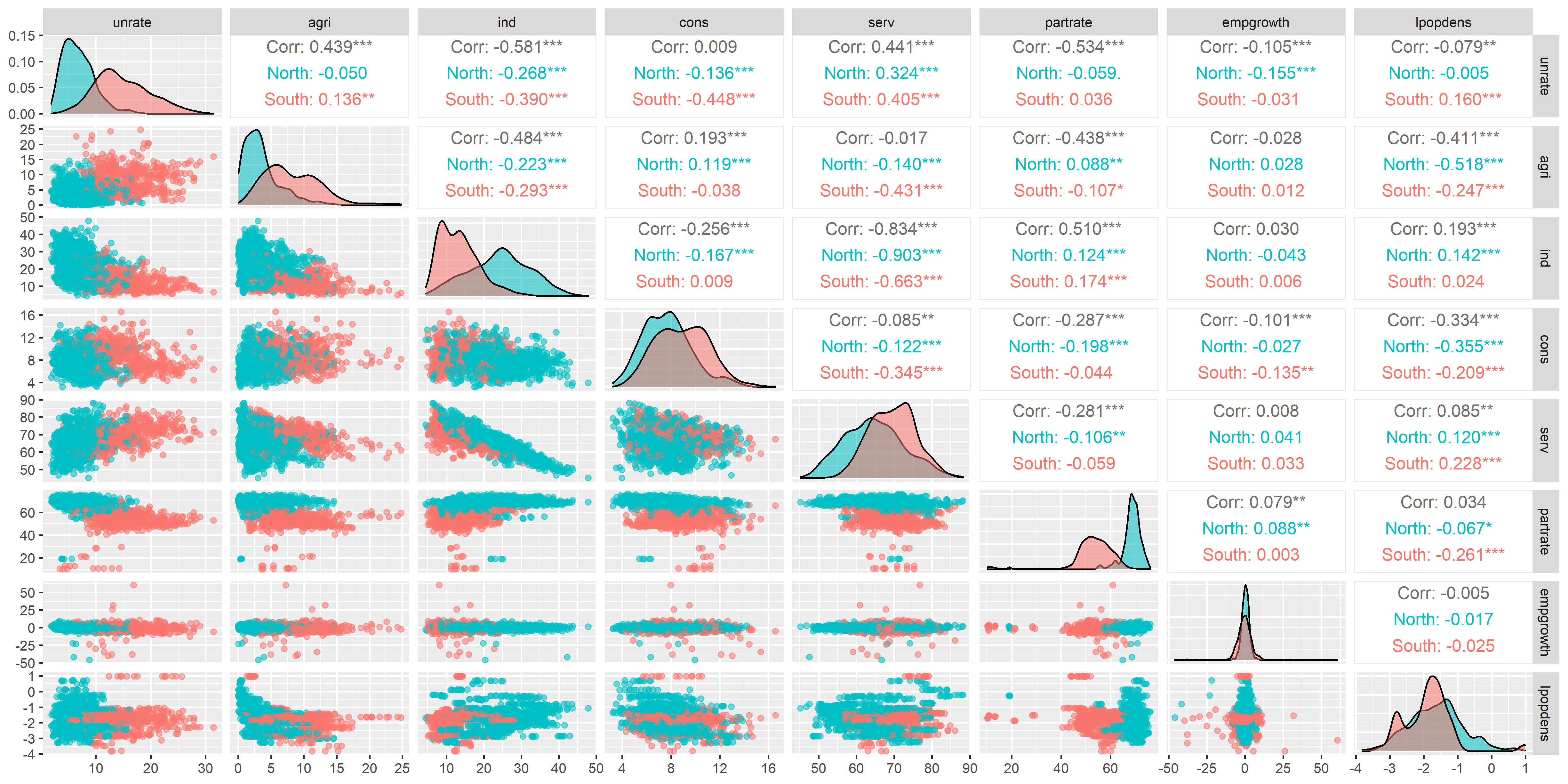} 
	\end{center}
	\caption{Scatter plots, histograms and correlations of pairs of variables separated into North and South.}
		\label{fig:ggpair} 
\end{figure}

Figure~\ref{fig:ggpair} shows scatter plots, histograms, and Pearson correlations of the unemployment rate and the seven covariates, grouped into northern and southern provinces. Additionally, the Pearson correlation coefficient is displayed for all pairs as well as for their northern and southern parts separately. The figure shows that, while variables like $unrate$, $agri$, $ind$ or $partrate$ exhibit the described dichotomy, there is no clear distinction between the North and the South for the remaining variables. We observe a negative correlation between the unemployment rate and $ind$, and a positive one with $serv$. The sign of the correlation can change when looking at the North and South separately, as in the case of $partrate$ or $agri$. Furthermore, we see a strong negative correlation between the employment in the industry and service sector. The response and explanatory variables are standardized separately by subtracting the overall sample mean and dividing by the overall sample standard deviation. Due to political reformation in Italy and the abolishment of seven provinces, missing data was imputed using the predictive mean matching method via the R package \texttt{mice} \citep{mice}.\footnote{As a robustness check and to avoid problems with compositional data, we additonally estimated the model on $\log$-ratios of $agri$, $ind$, and $serv$ against $cons$ as baseline economic sector. The results are comparable and show only minor qualitative differences. For this reason and in order to stay consistent with \cite{Minguez2020}, we continue our analysis with the original dataset.}

\section{Model specification}\label{sec:Model}
In this section, we introduce a novel Bayesian hierarchical model that allows for the study of spatio-temporal dependencies and uncovering the underlying clustering structure of areal data. The former is achieved by including random effects via a CAR prior and is described in Section~\ref{sec:likelihood}. The latter is achieved by using a Bayesian nonparametric prior for the areal-specific parameters. Section~\ref{sec:bnp} describes this feature and gives a brief overview of the DP.

\subsection{Likelihood and spatio-temporal random effects}
\label{sec:likelihood}
Let $\bm Y = \left[ Y_{it} \right]$, $i = 1, \dots, I$, $t = 1, \dots, T$, be a matrix of areal observations, such that $\bm Y \in \mathbb{R}^{I \times T}$. Each entry $Y_{it}$ represents the observed value of interest at the $i$-th areal unit at time $t$, which in the application under study corresponds to the unemployment rate in province $i$ at year $t$. For each areal unit and at each time point, we consider a set of $p$ predictors and encode them together with an intercept term in the $(p+1)$-dimensional column vector $\bm x_{it}$. For $i = 1, \dots, I$ and $t = 1, \dots, T$, we model $Y_{it}$ as follows:
\begin{align}\label{eq:Yit_regression}
	&Y_{it} = {\bm x_{it}}' \bm \beta_i + w_{it} + \epsilon_{it}, \quad \epsilon_{it} \iid \text{N}\left(0, \sigma^2\right) 
\end{align}
where ${\bm x_{it}}'$ represents the transpose of the vector ${\bm x_{it}}$, $\bm \beta_i$ the vector of regression parameters for area $i$, and $w_{it}$ a spatio-temporal random effect, discussed later in detail. Furthermore, we assume areal-specific Gaussian error terms with variance $\sigma^2$.

Let $\bm X_t = \left[ \bm x_{1t}, \bm x_{2t},  \ldots, \bm x_{It} \right]'$ be the matrix of predictors at time $t$ and let $\bm w_t = \left(w_{1t}, \dots, w_{It} \right)'$ and $\bm \epsilon_t = \left(\epsilon_{1t}, \dots, \epsilon_{It} \right)'$ be the vectors of random effects and error terms, respectively. Moreover, let $B = \left[\bm \beta_1, \bm \beta_2, \ldots, \bm \beta_I\right]$ be a matrix containing the regression coefficients.
Eq.~\eqref{eq:Yit_regression} can then be re-written in vector form, for each $t = 1, \dots, T$, as follows:
\begin{equation}\label{eq:Yt_regression}
	\bm Y_t = diag\left(\bm X_t B\right)  +  \bm w_t + \bm \epsilon_t
\end{equation}
where $diag\left(\bm X_t B\right)$ indicates the main diagonal of the matrix $\bm X_t B$ and $\bm Y_t$ the observations for all areal units at time $t$. The notation adopted in Eq.~\eqref{eq:Yt_regression} is convenient, as it allows modeling of the spatio-temporal random effects at time $t$ directly. In this work, we opt for an autoregressive decomposition similar to the one described by \cite{RUSHWORTH201429} and \cite{lee2016}. The temporal correlation is induced through the conditional expected value, while the precision matrix induces the spatial correlation. Specifically:
\begin{align}\label{eq:ARCAR}
	&\bm w_t|\bm w_{t-1} \sim \text{N}_I\left(diag\left(\bm \xi\right) \bm w_{t-1}, \tau^2 Q\left(\rho, W\right)^{-1} \right), \quad t = 2, \dots, T \\
	&\bm w_1 \sim \text{N}_I\left(\bm 0, \tau^2 Q\left(\rho, W\right)^{-1}\right) \nonumber
\end{align}
where $\bm \xi = (\xi_1, \dots, \xi_I)$ is a vector of autoregressive coefficients and $diag\left(\bm \xi\right)$ is a diagonal matrix with $\bm \xi$ as its entries. The random effects in Eq.~\eqref{eq:ARCAR} are modeled as a vector autoregressive process of order one, in short VAR(1). The covariance matrix is composed of the scale parameter $\tau^2$ and the inverse of the matrix $Q\left(\rho, W\right)$.

From Eq.~\eqref{eq:ARCAR} it is clear that the choice of $Q\left(\rho, W\right)$ plays a crucial role in modeling the spatial correlation. The matrix $Q\left(\rho, W\right)$ is of dimension $I \times I$ and depends on two quantities, namely the scalar parameter $\rho$ and the neighboring matrix $W$. The neighboring matrix $W \in \{0,1\}^{I \times I}$ is application-specific and reflects the contiguity structure of the $I$ areal units. Specifically, $W_{i,j} = 1$ if areal units $i$ and $j$ are adjacent (i.e., they are neighbors), and $W_{i,j} = 0$ otherwise. Following \cite{Leroux2000}, we define $Q\left(\rho, W\right) = \rho\left(diag\left(W \bm 1\right) - W\right) + \left(1-\rho\right) \mathbb I_I$, where $\mathbb{I}_I$ is the $I$-dimensional identity matrix and $\bm 1$ is a $T$-dimensional vector of ones. Let $v_i$ be the number of neighbors of site $i$. The matrix $diag\left(W \bm 1\right) - W$ has elements equal to $v_i$ if $i=j$, equal to $-1$ if $i$ and $j$ are neighbors, and equal to $0$ otherwise. The parameter $\rho$ allows modeling the spatial correlation among the random effects: 
$\rho=0$ corresponds to independent spatial random effects, while $\rho=1$ corresponds to the ICAR prior. In the latter case, the expectation for the random effect in areal unit $i$ at time $t$, namely $w_{it}$, conditionally on the previous times and the other areal units, is given by the sum of $\xi_i w_{i t-1}$ and the average of the random effects in geographically adjacent areal units at time $t$. It is important to point out that the parameter $\rho$ itself is not a correlation parameter, and its precise interpretation can be difficult. While \cite{lee2016} choose to fix $\rho=1$ to enforce spatial smoothing, we assume that $\rho$ follows a beta distribution a priori. The choice of the hyper-parameters of this distribution is specified in Section~\ref{sec:prior_elicit}.

The joint distribution of $\tilde{\bm w}  = \left[\bm w_1,\ldots , \bm w_T \right]$ is a GMRF. Specifically, it has mean zero and its precision matrix
$\Omega$ is tri-block diagonal with $T$ blocks of dimension $I\times I$. See Section~\ref{sec:fullcond_randeff} for details on the derivation of $\Omega$. This representation enables the use of known algorithms for posterior sampling of the time varying effects $\bm w_{t}$. In particular, we implement the algorithm proposed in \cite{Mccausland2011}. The efficiency of the algorithm for banded matrices increases with lower bandwidths. As the precision matrix $Q$ inherits the bandwidth from $W$, we capitalize on this efficiency gain by first reorganizing $W$ into a banded matrix and then minimizing its bandwidth using the algorithm of \cite{cuthillmckee}.
The choice of the hyperparameters is specified in Section~\ref{sec:prior_elicit}.
\subsection{Bayesian nonparametric areal clustering}
\label{sec:bnp}
As mentioned in Section~\ref{sec:intro}, we are interested in detecting which areal units exhibit similar patterns. To this end, we allow for clustering through the inclusion of the DP prior for some of the areal-specific parameters. A process $P$ distributed as a DP can be seen as an infinite mixture of point-masses at i.i.d. locations:
\begin{equation*}
	P(\cdot) = \sum_{j = 1}^{+\infty}\omega_j \delta_{\theta_j}(\cdot)
\end{equation*}
where $\delta_{\theta}(x)$ is the Dirac's delta, equal to 1 when $x = \theta$, and zero otherwise. The other elements specifying the mixture are the infinite sequence of locations $\theta_1, \theta_2, \dots \iid P_0$ and the infinite sequence of weights, which follows the stick-breaking construction \citep{Sethuraman_94}:
\begin{align*}
	&\omega_j = v_j \prod_{l < j}(1 - v_l), \ j = 2, 3, \ldots, \quad \omega_1 = v_1\\
	&v_1, v_2, \dots \iid \text{Beta}(1, \alpha)
\end{align*}
with $\text{Beta}(a, b)$ representing the beta distribution with mean $a/(a + b)$ and variance $ab/((a + b)^2 (a + b + 1))$. The distribution $P_0$ is the base measure and represents the mean distribution around which the DP is centered, while the mass parameter $\alpha > 0$ indicates its variability around $P_0$. This definition of the DP highlights the discreetness of its trajectories, which in turn implicitly induces clustering. 
Letting $\bm \phi_i = \left(\bm \beta_i, \xi_i\right) \in \mathbb R^{p + 2}$,  $i = 1, \dots, I$,
and placing a DP prior on the vectors $\bm \phi_1, \dots, \bm \phi_I$ allows for clustering the regression coefficients $\bm \beta_i$ as well as the temporal persistencies $ \xi_i$. We write:
\begin{align*}
	&\bm \phi_1, \dots, \bm \phi_I | P \iid P \\
	&P \sim DP(\alpha, P_0) \nonumber
\end{align*}
This can be equivalently represented through the introduction of a vector of areal-specific allocation variables $\bm s = (s_1, \dots, s_I)'$ and a set of unique values $\bm \phi^\star = \left(\bm \phi^\star_1, \dots, \bm \phi^\star_{K_I}\right)$, with $K_I \leq I$ and such that $s_i = j \iff \bm \phi_i = \bm \phi^\star_j$. This representation defines a partition $\{C_1, C_2, \ldots C_{K_I} \}$ composed of $K_I$ clusters $C_j = \{i \in \{1, \dots, I\} | s_i = j\}$, for $j = 1, \dots, K_I$. To each cluster corresponds a distinct value of the parameter vector $\bm \phi^\star_j$. The vector of allocations $\bm s$ follows, a priori, a P\'{o}lya urn scheme with allocation probabilities:
\begin{align*}
&\text{P\'{o}lyaUrn}(\bm s \mid \alpha) = P(s_1)\prod_{i = 2}^I P\left(s_i \mid s_1, \ldots, s_{i-1}\right) \\
& P\left(s_i \mid s_1, \ldots, s_{i-1}\right) = 
\begin{cases}
	&\dfrac{n_{ij}}{i-1 + \alpha}, \quad j = 1, \dots, K_I\\
	&\\
	&\dfrac{\alpha}{i-1 + \alpha}, \quad j = K_I+1
\end{cases} 
\end{align*}
where $n_{ij}$ is the size of the $j$-th cluster before we assign the $i$-th observation, that is the size of $\{i' < i | s_{i'} = j\}$. This means that the joint prior of $\bm s$ can be computed as the product of all conditional distributions of $s_i$, given $s_1,\ldots,s_{i-1}$, for all $i=1,\ldots,I$. The value assumed by $s_i$ in this conditional distribution is either an ``old'' value among already observed $s_1,\ldots,s_{i-1}$ or a ``new'' value, corresponding to an additional cluster. The first item is always allocated to cluster 1, i.e. $P(s_1=1)=1$, and it is associated to the first unique value $\bm \phi^\star_1$. Let us define $\bm \xi^{\star}_{\bm s} = \left(\xi^\star_{s_1}, \dots, \xi^\star_{s_I} \right)$ and $B_{\bm s}^\star = \left(\bm \beta^\star_{s_1}, \dots,\bm \beta^\star_{s_I} \right)$. Conditioning on the allocation variables $\bm s$ and specifying priors for the remaining parameters, we can write the full model as follows, for all $i=1,\ldots,N$ and $t=1,\ldots,T$:
\begin{align}\label{eq:full_model}
	&Y_{it} \mid \bm x_{it}, \bm \beta^\star_{s_i}, w_{it}, \sigma^{2}, s_i \stackrel{ind} \sim \text{N}\left(\bm x_{it}' \bm \beta^\star_{s_i} +  w_{it}, \sigma^{2}\right) \nonumber \\
	&\bm w_t \mid \bm w_{t-1}, \bm \xi^\star_{\bm s}, \bm s, \tau^2, \rho, W \sim \text{N}_I\left( diag(\bm \xi^\star_{\bm s}) \bm w_{t-1}, \tau^2 Q( \rho , W)^{-1} \right) \nonumber \\
	&\bm w_1 \mid \tau^2, \rho, W \sim \text{N}_I\left(\bm 0, \tau^2 Q(\rho, W)^{-1}\right) \\
	& \sigma^2 \sim \text{Inv-Gamma}\left(a_{\sigma^2}, b_{\sigma^2}\right) \nonumber \\
	& \tau^2 \sim \text{Inv-Gamma}\left(a_{\tau^2}, b_{\tau^2}\right) \nonumber \\
	& \rho \sim  \text{Beta}\left(\alpha_{\rho} , \beta_{\rho} \right)\\
	& \bm s \mid \alpha \sim \text{P\'{o}lyaUrn}(\bm s \mid \alpha) \nonumber \\
	&\alpha \sim \text{Gamma}\left(a_{\alpha}, b_{\alpha}\right) \nonumber \\
	&\bm \phi^\star_1, \dots, \bm \phi^\star_{K_I} | \bm \mu_{\bm \beta}, \Sigma_{\bm \beta}, a_{\xi}, b_{\xi} \iid P_0, \quad \bm \phi^\star_j = \left(\bm \beta^\star_j, \xi^\star_j\right), \quad j = 1, \dots, K_I \nonumber \\
	&P_0\left(d\bm \phi^\star\right) = \text{N}_{p + 1}\left(d\bm \beta^\star \mid \bm \mu_0, \Sigma_0\right) \text{Beta}_{(-1,1)}\left(d\xi^\star \mid a_{\xi}. b_{\xi}\right) \nonumber
\end{align}
which completes the model description by specifying the prior distribution for the location parameters, $P_0(d\bm \phi^\star)$, and for the hyperparameters  $\sigma^2$, $\tau^2$, and $\alpha$. In Eq.~\eqref{eq:full_model},
\text{Beta}$_{(-1,1)}(a,b)$ stands for the transformed beta distribution over the interval $(-1,1)$ with parameters $a, b > 0$, obtained by applying a linear transformation $2B-1$ to a standard $\text{Beta}_{(0,1)}(a,b)$-distributed random variable $B$. We denote by $\text{Inv-Gamma}\left(a, b\right)$ the inverse-gamma distribution with mean $b/(a-1)$ and mode $b/(a+1)$, where $a$ and $b$ are the shape and scale parameter respectively. Moreover, we denote with 
$\text{Gamma}\left(a, b\right)$ the gamma distribution with shape parameter $a$ and rate parameter $b$. 

Note that, since we are clustering the areal-specific time series $\bm Y_i$, $i=1, \ldots, I$, the proposed model implies a clustering of time series. Cluster estimates of the areal units are based on the posterior distribution of the vector of allocation variables $\bm{s}$, obtained minimizing the posterior expectation of a suitable loss function. Some of the most popular choices of loss functions for partitions include the Binder loss function \citep{binder}, the variation of information \citep[VI,][]{vi, wade2018bayesian}, or generalizations thereof \citep{salso-paper}. Since the estimated number of clusters largely depends on the chosen method of loss calculation, we conduct an extensive sensitivity analysis that can be found in Appendix~\ref{sec:sens_loss}. The flexibility of Bayesian hierarchical models allows for a multitude of similar alternative model specifications, especially regarding the choice of the prior. To this end, we compare possible alternatives and present the results in Appendix~\ref{sec:alt_model_spec}.
\section{Algorithm}
\label{sec:algorithm}

Let $\bm \vartheta$ be the vector of unknown model parameters, that is $\bm \vartheta= (\bm \beta^\star, \sigma^{2}, \bm \xi^\star, \bm s, \bm w_1, \ldots,  {\bm w}_T, \tau^2, \rho, \alpha)$.
To obtain posterior samples from the joint distribution of the parameters, we implement a Metropolis-Hastings within Gibbs sampler. After setting initial values for the parameters  $\bm \beta^\star, \sigma^{2}, \bm \xi^\star, \bm s, \bm w_1, \ldots, \bm w_T$, $\tau^2$, $\rho$, and $\alpha$, at each iteration of the MCMC algorithm we perform the following steps:
\begin{enumerate}
	\item Sample the allocation variables $s_i$, $i = 1, ..., I$, through the algorithm described in \cite{FavTeh13}, which in our case amounts to an extension of the popular Algorithm 8 by \cite{Neal00}, that includes a re-use step. Details can be found in Appendix~\ref{sec:fullcond_bnp_remain}.
	
	\item Sample the unique values $(\bm \beta^\star, \bm \xi^\star)$ conditioned on all the other parameters and the allocation variables $\bm s$ . The full conditional distribution for $\bm \beta^\star$ is conjugate and Gaussian, while sampling $\bm \xi^\star$ requires a Metropolis-Hastings step. Further details can be found in Appendix~\ref{sec:fullcond_bnp_remain}.
	
	\item Sample $\alpha$ from a mixture of gamma distributions using the auxiliary variable sampler by \cite{west1992hyperparameter}. Details can be found in Appendix~\ref{sec:fullcond_bnp_remain}.
	
	\item Sample the spatio-temporal random effects $( {\bm w}_1, \dots,  {\bm w}_{T})'$ from a multivariate normal distribution using the algorithm proposed in \cite{Mccausland2011}. Details can be found in Appendix~\ref{sec:st_randeff}.
	
	\item Sample $\sigma^2$ from an inverse-gamma distribution. Details can be found in Appendix~\ref{sec:fullcond_bnp_remain}.

	\item Sample $\tau^2$ from an inverse-gamma distribution. Details can be found in Appendix~\ref{sec:fullcond_bnp_remain}.
	
	\item Sample $\rho$ using a Metropolis-Hastings step, specifically a random walk on the logit transformation of $\rho$. Details can be found in Appendix~\ref{sec:fullcond_bnp_remain}.

\end{enumerate}

\section{Empirical application}
\label{sec:emp_app}


In the following, we analyze the Italian unemployment data using the proposed model and algorithm introduced in Sections~\ref{sec:Model} and \ref{sec:algorithm}, respectively. We carefully elicit the prior distributions used in the model, discuss strategies to choose a representative partition and interpret the implied economic findings. To conclude, we benchmark the proposed approach in terms of out-of-sample forecasting capabilities against a variety of competitor models.

\subsection{Prior elicitation and sampling details}
\label{sec:prior_elicit}
We set the hyperparameters of $P_0$ to be $\bm{\mu}_0=\bm{0}$, $\Sigma_0=\mathbb{I}_{p+1}$, $\alpha_{\xi}=1$, $\beta_{\xi}=1$, representing weak prior information. For $\sigma^2$ and $\tau^2$ we choose an inverse-gamma distribution with parameters $\alpha_{\sigma^2}=\alpha_{\tau^2}=3$ and $\beta_{\sigma^2}=\beta_{\tau^2}=2$, implying unitary a-priori mean and variance for both parameters. In line with the literature \citep[e.g.,][]{Banerjee2014}, we set an informative prior for $\rho$ via a beta distribution with $\alpha_{\rho}=6$ and $\beta_{\rho}=1$, so that $\mathbb{E}[\rho] = 6/7$ and $\mathbb{V}[\rho] \approx 0.015$.

Due to its influential nature on the posterior, in particular on the number of clusters, we do not fix the mass parameter $\alpha$ but instead assume that a priori it follows gamma distribution with parameters $a_{\alpha}=3$ and $b_{\alpha}=2$, yielding $\mathbb{E}[K_I] \approx 6.75$ and $\mathbb{V}[K_I] \approx 7.32$ . These numbers are in agreement with prior information on the number of groups of provinces where the economic development of Italian regions is different, e.g. north-west, north-east, center, south, and the two islands.

\begin{figure}[t]
	\centering
	\includegraphics[trim = 400 0 400 0, clip = true, width=.7\textwidth]{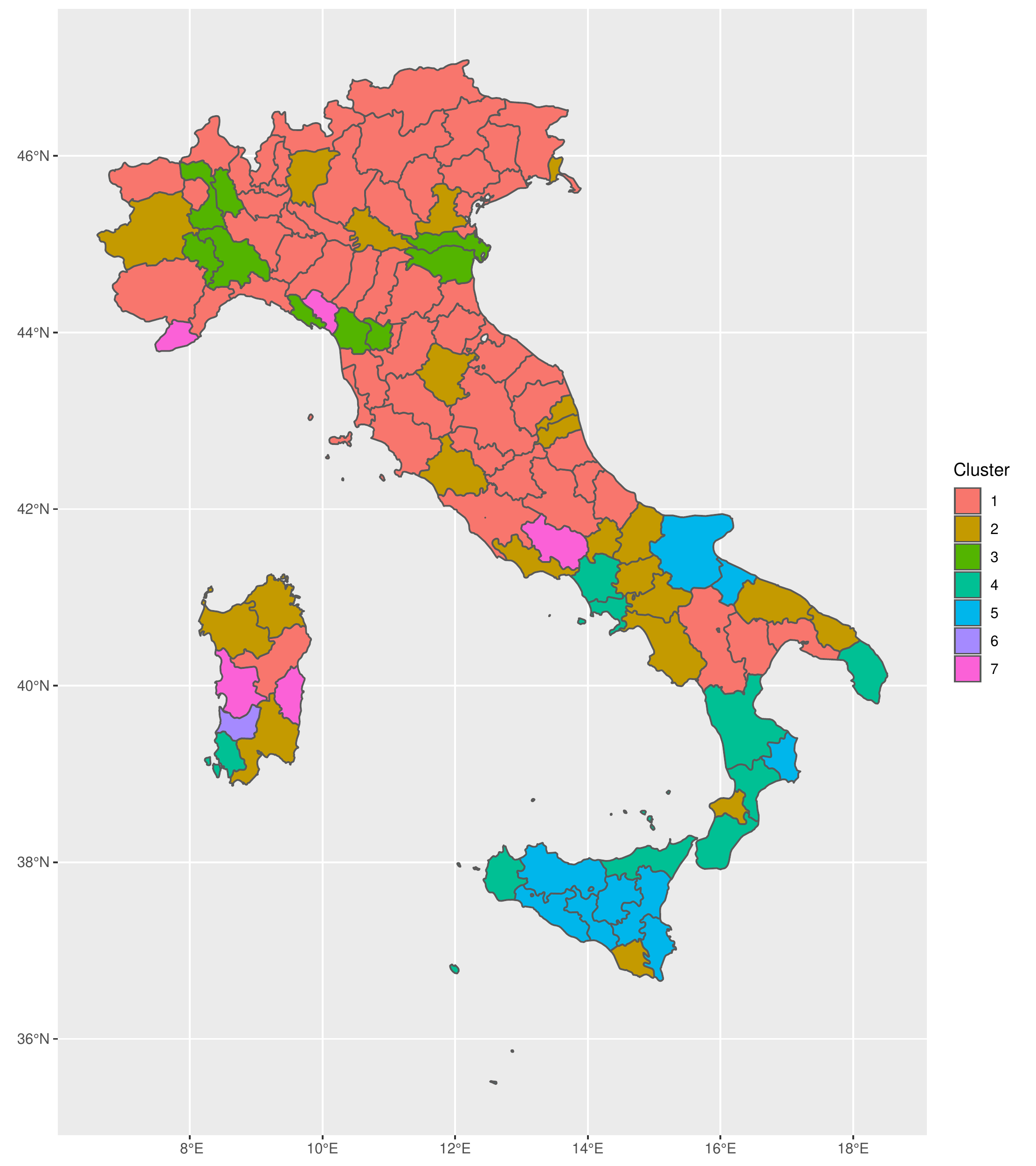}
	\caption{Partition of the Italian provinces obtained by minimizing the posterior expectation of the Binder loss function. A detailed sensitivity analysis motivating the choice of the method of loss calculation can be found in Appendix~\ref{sec:sens_loss}. }
	\label{fig:italy_clust}
\end{figure}

The use of a Bayesian nonparametric prior in the proposed model allows for clustering of the areal units (i.e., the Italian provinces). To provide an point estimate of the partition of the provinces, we minimize the posterior expectation of specific loss functions, namely the Binder loss \citep{binder} and the Variation of Information \citep[VI,][]{wade2018bayesian}. 
These methods of expected loss calculation are invariant to the label-switching problem and can therefore be applied directly to the posterior samples of $\bm s$.An extensive comparison of these methods is presented in Appendix~\ref{sec:sens_loss}. In what follows, we present details based on the chosen partition estimated by minimizing the posterior expectation of the Binder loss function.
The comparison metrics in Appendix~\ref{sec:sens_loss} confirm that the model retains predictive accuracy under the chosen partition, while the lower number of clusters of mostly contiguous provinces allows for a clearer economic analysis. The partition is displayed in Figure \ref{fig:italy_clust}.

\subsection{Posterior Inference}\label{sec:econint}
We continue by analyzing the clustering in Figure \ref{fig:italy_clust} and its implications on the underlying unemployment differentials. To this aim, we re-run the MCMC algorithm conditionally on the Binder loss partition estimate and summarize the estimated $\bm \beta$ coefficients within each cluster. 
We discard 10,000 draws as burn-in and keep every 3rd draw of the remaining 15,000 draws, thus using the 5,000 remaining draws for posterior inference. The posterior means of the covariate effects are reported in Table \ref{tab:res1}.
\begin{table}[t]
\centering
	\caption{Posterior means of the regression coefficients $\bm \beta$ within each cluster of the estimated, selected partition.}
		\label{tab:res1}
		\begin{tabular}{lcccccccc}
			\hline
			\hline
		Cluster (size) & intercept  & agri & ind & cons & serv  & partrate & empgrowth & lpopdens \\ 	
			\hline
            Cluster 1 (55) & -0.282 & -0.402 & -0.947 & -0.201 & -0.784 & -0.090 & -0.056 & 0.057 \\ 
            Cluster 2 (22) & 0.056 & 0.433 & 0.620 & 0.157 & 0.755 & 0.022 & -0.030 & -0.002 \\ 
            Cluster 3 (9) & -0.536 & -0.150 & -0.092 & 0.036 & -0.050 & 0.764 & -0.212 & 0.114 \\ 
            Cluster 4 (9) & 1.924 & 0.042 & 0.590 & -0.226 & 0.459 & 0.633 & -0.171 & 0.130 \\ 
            Cluster 5 (9) & 2.786 & 0.119 & 0.299 & 0.067 & 0.318 & 1.034 & -0.166 & -0.251 \\ 
            Cluster 6 (1) & 1.268 & -1.710 & -0.108 & 0.227 & 0.491 & -0.064 & -0.314 & -1.448 \\ 
            Cluster 7 (5) & 0.481 & -0.687 & -1.151 & -0.344 & -1.021 & 0.658 & -0.163 & -0.319 \\ 
			\hline	
		\end{tabular}
\end{table}


Figure \ref{fig:italy_clust} shows that Cluster 1 is predominantly made up of northern and central provinces, constituting the largest cluster. Interestingly, the province Taranto as well as the southern regions Abruzzo and Basilicata are also assigned to this cluster. A possible reason could be that, amongst all southern provinces, these have strikingly low unemployment rates.
The second cluster contains provinces which are scattered geographically and three out of eight provinces of Sardinia. The third cluster is made up of a small group of northern provinces like for example Alessandria, Vercelli, and Novara, while the fourth contains a few southern provinces: Carbonia-Iglesias in Sardinia, Napoli, Caserta, Lecce, in Calabria the provinces Cosenza, Catanzaro, and Reggio Calabria, and Messina, and Trapani in Sicily. Cluster 5 contains almost all of Sicily, Foggia, Barletta-Andria-Trani, and Crotone. Cluster 6 is a singleton cluster made up of the province Medio Campidano. Its classification as a single cluster is consistent among most partitions examined in Appendix~\ref{sec:sens_loss} and is attributable to the highly negative coefficients of $agri$ and $lpopdens$. Cluster 7 contains the provinces Imperia, Massa Carrara, Frosinone, Oristano, and Ogliastra. We can observe in  Figure~\ref{fig:choropleth} that they exhibit similar average unemployment rates. Note that it is the only cluster where the coefficients corresponding to the economic sectors are all significantly negative.

\begin{figure}[p]
	\centering
	\includegraphics[width=.93\textwidth]{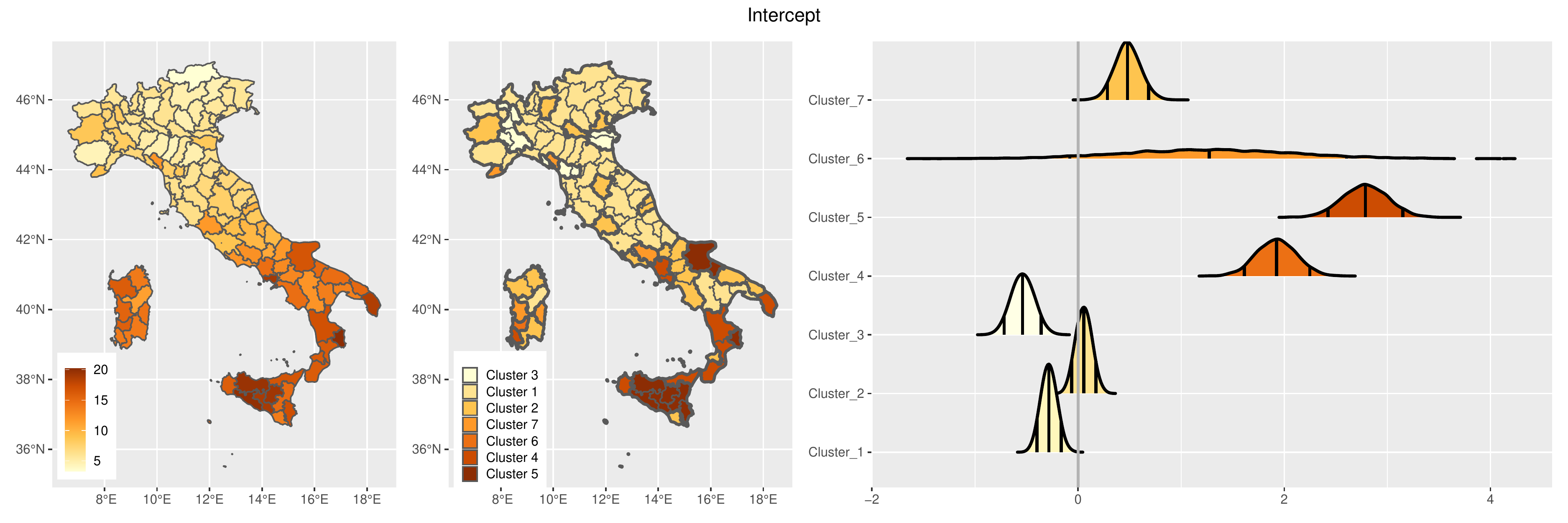}\\
	\includegraphics[width=.93\textwidth]{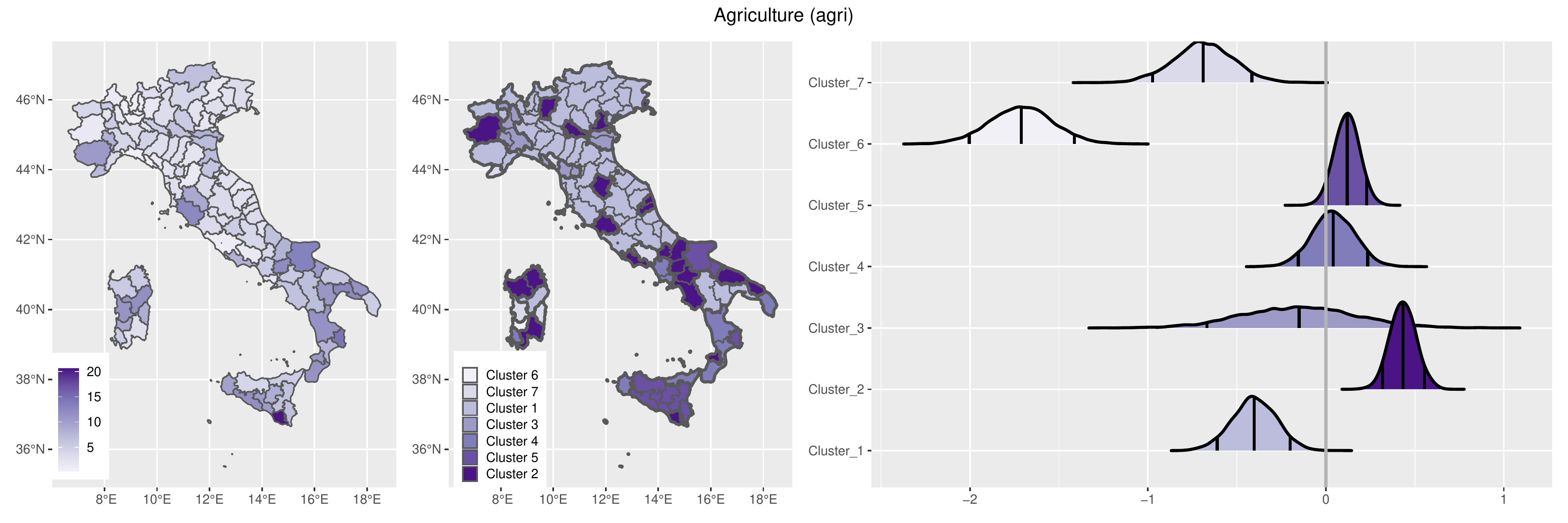}\\
	\includegraphics[width=.93\textwidth]{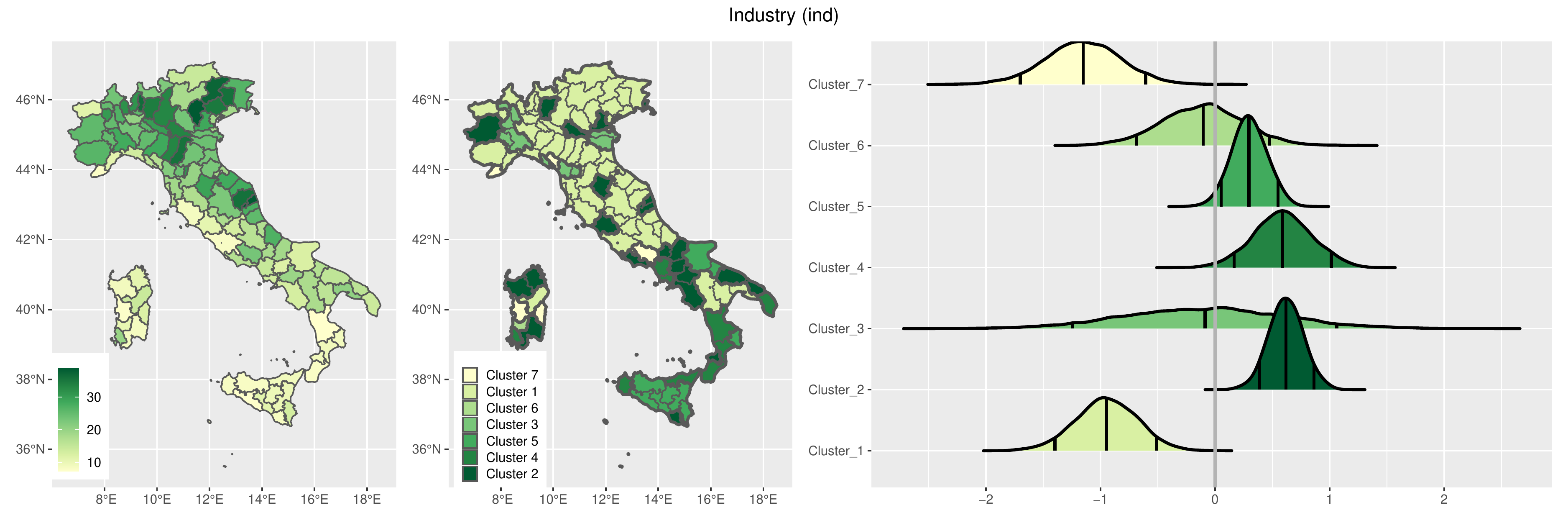}\\
	\includegraphics[width=.93\textwidth]{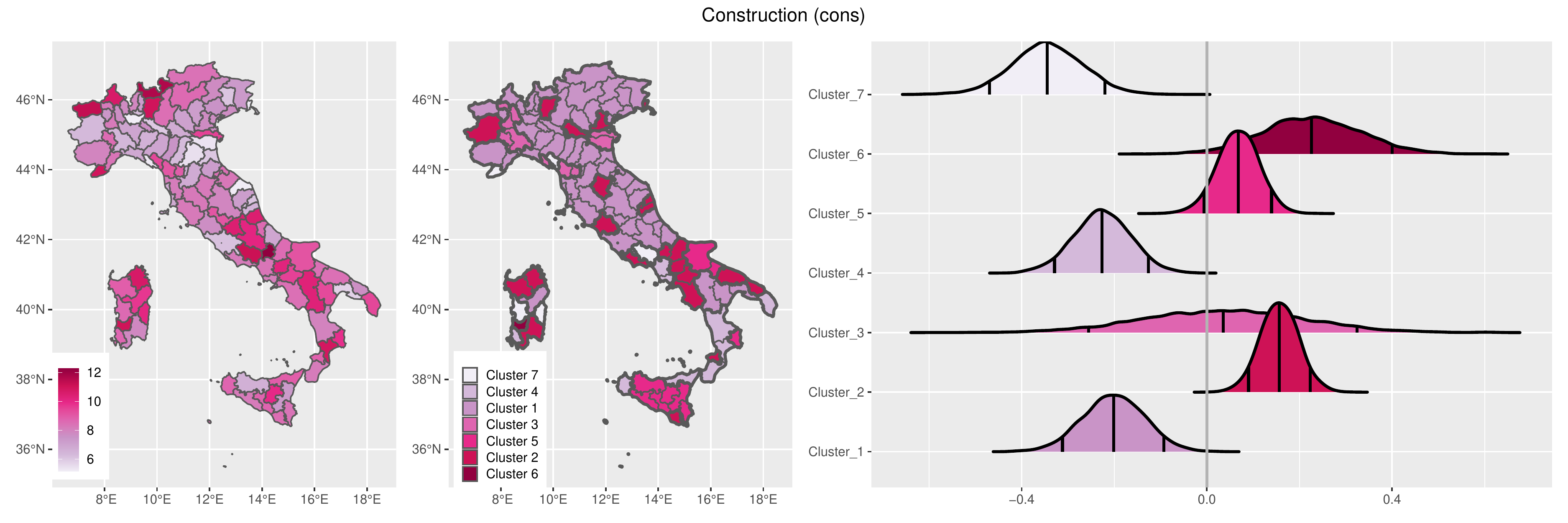}
	\caption{Visualization of $intercept$, $agri$, $ind$, $cons$ (top to bottom). Unstandardized covariates, averaged over time (left). Posterior means (middle) and posterior kernel density estimates with $5\%$, $50\%$, and $95\%$ quantiles (right) of the corresponding regression coefficients. The first map (top-left) shows the average unstandardized response.}
	\label{fig:xb_maps1}
\end{figure}

\begin{figure}[p]
	\centering
	\includegraphics[width=.93\textwidth]{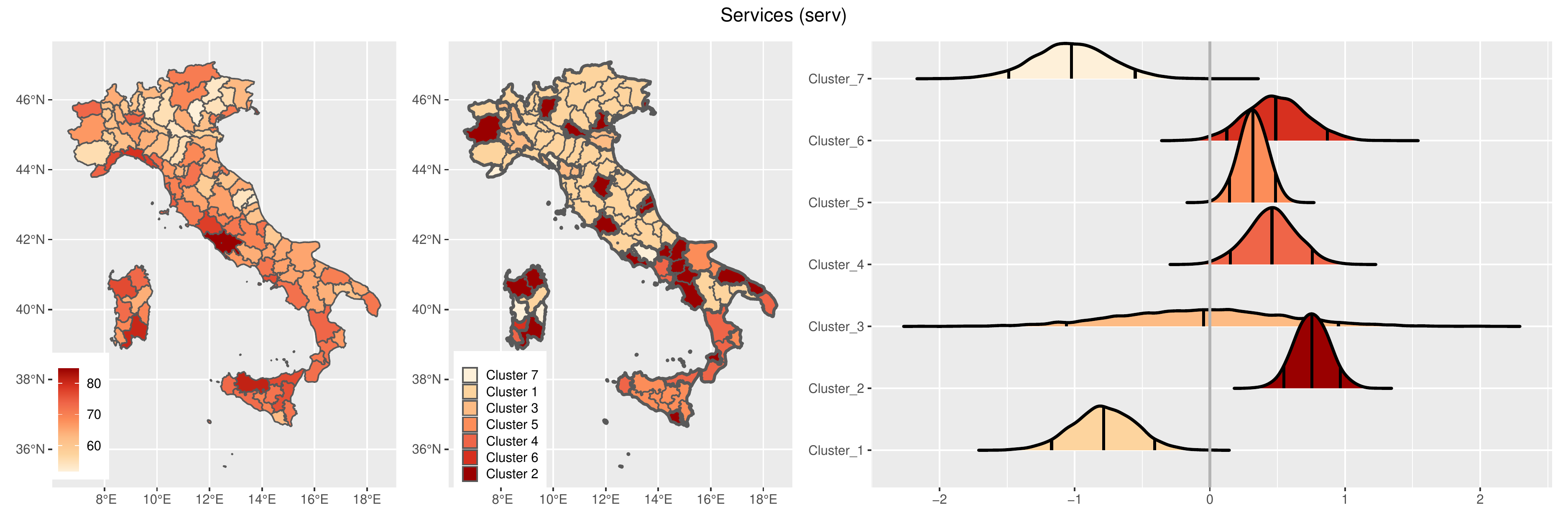}\\
	\includegraphics[width=.93\textwidth]{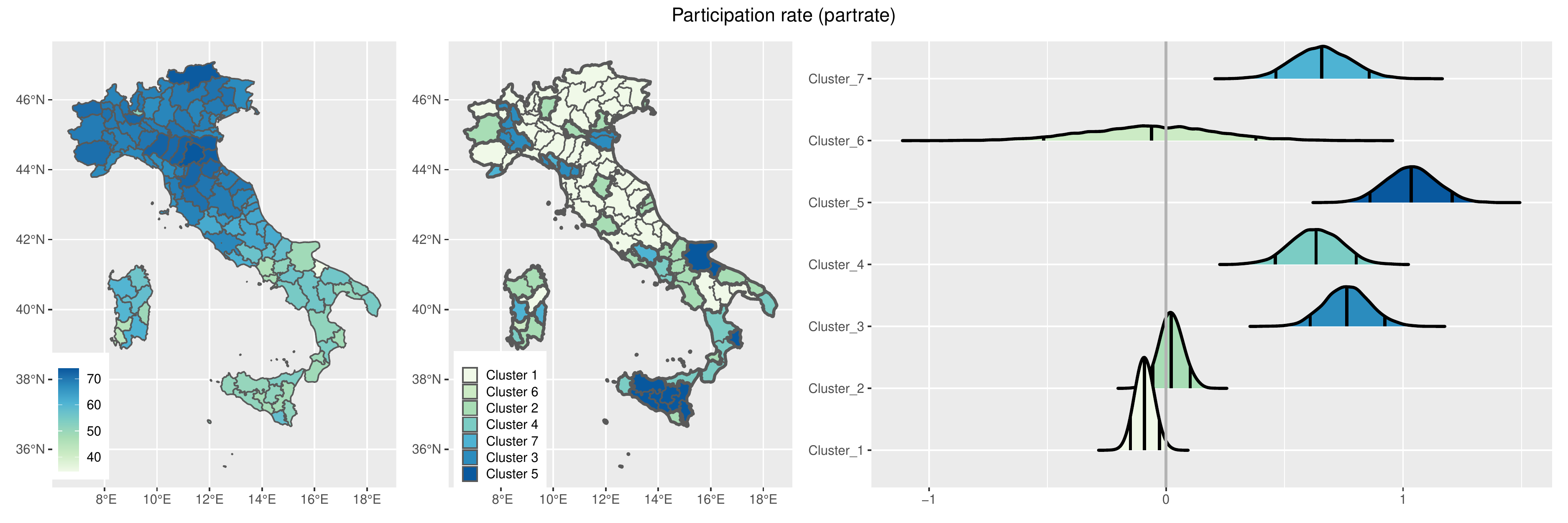}\\
	\includegraphics[width=.93\textwidth]{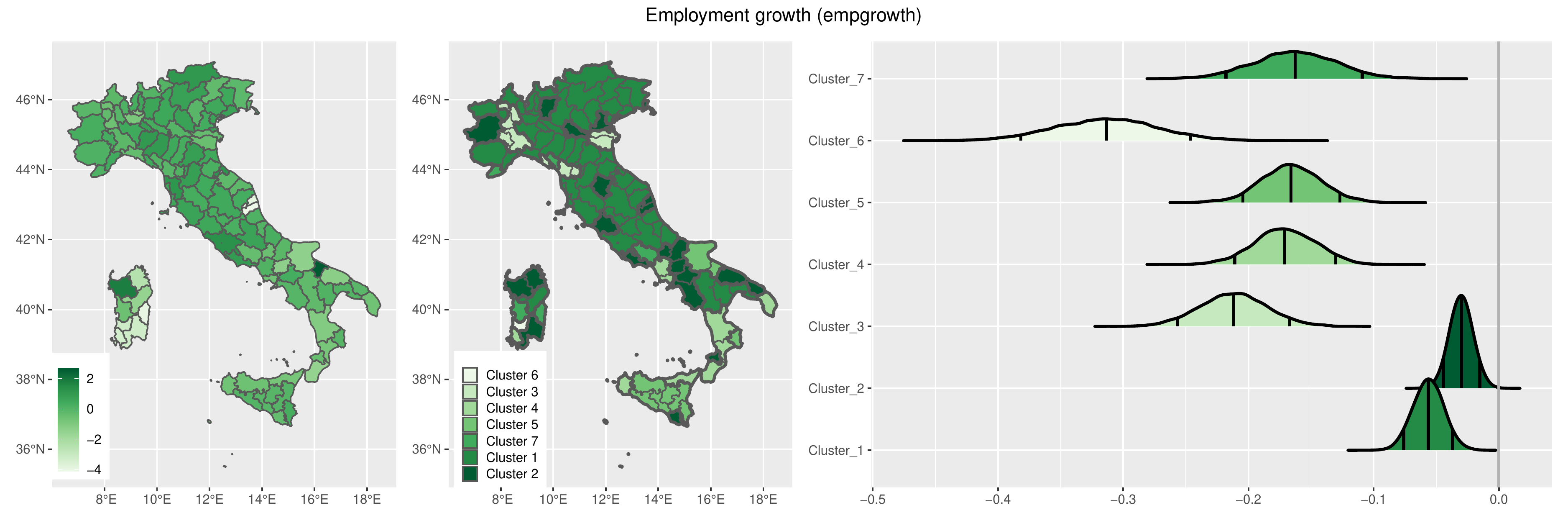}\\
	\includegraphics[width=.93\textwidth]{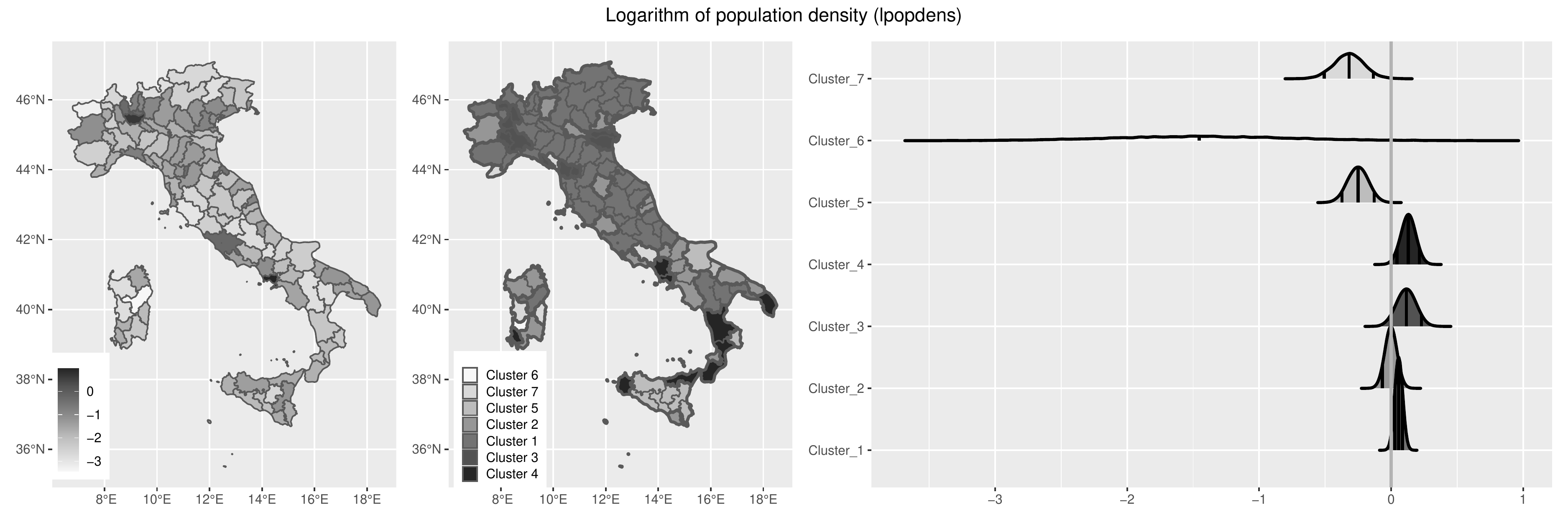}
	\caption{Visualization of $serv$, $partrate$, $empgrowth$, $lpopdens$ (top to bottom). Unstandardized covariates, averaged over time (left). Posterior means (middle) and posterior kernel density estimates with $5\%$, $50\%$, and $95\%$ quantiles (right) of the corresponding regression coefficients.}
	\label{fig:xb_maps2}
\end{figure}

One of the most pronounced results are the positive and negative intercepts for the southern and northern provinces, respectively. This result clearly captures the well-known North-South dichotomy, also displayed in Figure~\ref{fig:choropleth}. 
In order to facilitate the interpretation of the cluster-specific coefficients, Figures~\ref{fig:xb_maps1} and \ref{fig:xb_maps2} depict the average unstandardized observed covariates, the corresponding posterior mean $\hat{\bm \beta}$ and the kernel density estimates of the posterior distribution of $\bm \beta$. As every cluster has only one set of $\bm \beta$ coefficients, each of the maps depicting the estimated values shows seven different hues of the chosen color.

Observing both a negative and a positive effect of the participation rate might seem ambiguous, but is in line with the literature according to \cite{elhorst_mystery}. A negative effect can stem from the fact that low participation rates are often coupled with low investments in human capital and low commitment to work, both attributes spurring unemployment. The positive effect on the other hand could indicate an insufficient job offer.

Cluster 2 includes provinces with a somewhat puzzling variety of economic and social characteristics. It encompasses the wealthier Bergamo as well as Torino, where the social fabric of the metropolitan area is speckled, and some southern provinces. The latter, however, exhibit comparably low unemployment rates among the South. As a result, the estimates of the intercept and the $\bm \beta$ coefficients associated to  $partrate$, $lpopdend$, and $empgrowth$ are close to zero, as can be seen in Table~\ref{tab:res1} and Figures~\ref{fig:xb_maps1} and \ref{fig:xb_maps2}.
Attributes that distinguish Cluster 3 from the rest of the North are most notably the comparably high coefficients of $partrate$ as well as the low coefficient of $empgrowth$. The maps in Figure~\ref{fig:xb_maps2}, corresponding to the latter, show that the provinces included in Cluster 3 have moderate values of employment growth and a very low $\bm \beta$ estimate, clearly standing out among the surrounding provinces. These results, together with the high $\hat{\beta}_{partrate}$, suggest an insufficient offer in jobs in this cluster.
Member provinces of the southern Clusters 4 and 5 are characterized by similar social issues, in addition to  high values of the unemployment rates. However, Table~\ref{tab:res1} shows that the two clusters differ by the effect of the population density on the unemployment rate. The main driver of this distinction is possibly the province Napoli which, as we can observe in Figure~\ref{fig:xb_maps2}, has the highest population density among all provinces. In addition, Cluster 5 shows an overall higher level of unemployment.

A one to one comparison of the estimated effects with the results in \cite{Minguez2020} is not possible, as our model estimates an individual set of coefficients for every cluster and additionally includes an intercept term. However, Section~\ref{subsec:compmod} presents a comparison in terms of out-of-sample prediction.

\begin{figure}[t]
	\centering
	\hspace*{-1cm}
	{\includegraphics[width=16cm,height=8cm]{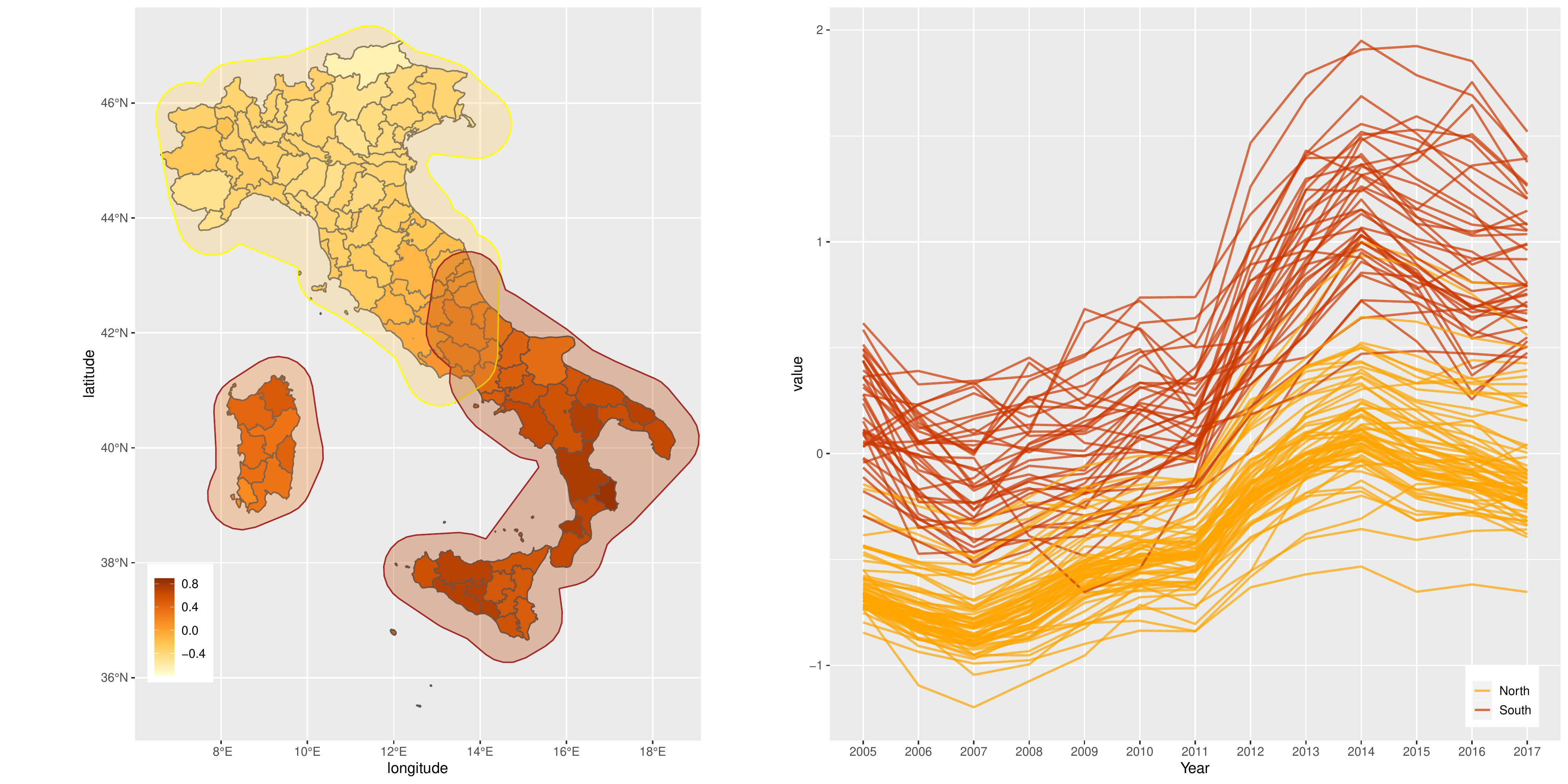}}%
	\caption[$\hat{\beta}_{popdens}$]{Average estimated spatio-temporal random effects (left panel) and the estimated time series for each province (right panel). The yellow and orange outlines on the map delineate the northern and southern Italian provinces, respectively.} 
	\label{fig:st_reff}
\end{figure}

The estimated spatio-temporal random effects are depicted in Figure~\ref{fig:st_reff}. Both the map and the time series show a clear distinction between northern and southern provinces. The map strongly resembles Figure~\ref{fig:choropleth}, implying that the random effects capture the differing levels of unemployment rates among the provinces. The pattern of the time series resembles Figure~\ref{fig:NorthS_2}, showing an upwards trend among all provinces. 

\subsection{Competitor models} \label{subsec:compmod}
In order to assess the performance and accuracy of the proposed model, we compare it to competitor models found in the literature on spatio-temporal data analysis. 
To ensure convergence and improve mixing for the proposed Bayesian spatio-temporal clustering model (BSTC), we run 25 independent MCMC chains, each initialized to different starting values. After discarding a burn-in of 5,000 iterations and keeping the remaining 4,000 draws per chain, we merge the results and obtain a final sample of 100,000 posterior draws.

In the following, we introduce the details for six increasingly elaborate frequentist models as well as the Bayesian spatio-temporal CAR model including a temporal autoregressive process (ST.CARar) proposed by \cite{CARBayesST}. The simplest model we use for comparison is a standard linear pooling model (Pooled), where 
the $(p+1) \times 1$ vector of coefficients $\bm{\beta}$ is restricted to be the same across time $t$ and space $i$:
\begin{align}
	Y_{it}= \bm{x}'_{it} \bm{\beta} + u_{it}, 	\quad u_{it} \sim N(0,\sigma^2) \label{eq:Pooled}
\end{align}
In order to model province-specific heterogeneity, we also consider splitting the error term $u_{i,t}$ into a province-specific fixed part alongside the i.i.d. random part, obtaining the individual fixed effects (IFE) model \citep{Baltagi_book}:
\begin{align}
	Y_{it}         =\bm{x}'_{it} \bm{\beta} + \tilde{u}_{it}, \quad
	\tilde{u}_{it} = \mu_i +v_{it}, \quad v_{it} \sim N(0,\tilde{\sigma}^2) \label{eq:IFE}
\end{align}
The estimation of the two models is performed via the R-package \texttt{plm} by \cite{plm_package}.

Both models can be augmented by a spatial-autoregressive (SAR) component $\rho \sum_{i \neq j} W_{ij} Y_{jt}$, where  $\rho$ is the spatial autocorrelation coefficient and $W_{ij}$ is the entry of $W$ corresponding to the areal units $i$ and $j$. Model \eqref{eq:Pooled} then becomes 
\begin{align}
	Y_{it}= \rho \sum_{i \neq j} W_{ij} Y_{jt} + \bm{x}'_{it} \bm{\beta} + u_{it}, \quad u_{it} \sim (0,\sigma^2) \label{eq:Pooled-SAR}
\end{align}
and is termed the Pooled-SAR model. Likewise, Model (\ref{eq:IFE}) becomes
\begin{align}
	Y_{it}= \rho \sum_{i \neq j} W_{ij} Y_{jt} + \bm{x}'_{it} \bm{\beta} + \tilde{u}_{it}, \quad \tilde{u}_{it} \sim N(\mu_i,\tilde{\sigma}^2) \label{eq:IFE-SAR}
\end{align}
and it is referred to as the IFE-SAR model. The augmented models \eqref{eq:Pooled-SAR} and \eqref{eq:IFE-SAR} are estimated via the R-package \texttt{splm} by \cite{splm_package}.

In order to incorporate smoothing across the spatial and temporal dimension simultaneously, \cite{Lee_2011_PSplineAnova} propose to explicitly model the interaction between space and time through some function $f$ depending on the spatial coordinates $s_1$ and $s_2$ and the temporal dimension $t$.
They develop an ANOVA method based on P-splines (PS-ANOVA) with the following model specification:
\begin{align}
	Y_{it}= f(s_{1i},s_{2i},t) + \bm{x}'_{it} \bm{\beta} + \epsilon_{it}
	\nonumber 
\end{align}
\cite{Minguez2020} augment the model with the spatial autoregressive component leading to the following specification of their PS-ANOVA spatial-autoregressive (PS-ANOVA-SAR) model:
\begin{align}
	Y_{it}= f(s_{1i},s_{2i},t) + \rho \sum_{j=1}^{N} W_{ij} Y_{jt} +
	\bm{x}'_{it} \bm{\beta} + \epsilon_{it}
	\nonumber 
\end{align}
The Bayesian competitor to the proposed model is the ST.CARar model implemented in the R package \texttt{CARBayesST} \citep{CARBayesST}. To facilitate comparison, we align the model specification with the proposed model by choosing Gaussian errors $\epsilon_{it}$ and the same prior distributions for $w_{it}$, $\tau^2$ and $\sigma^2$: 
\begin{align}
	&Y_{it} = \bm x'_{it} \bm \beta +  w_{it} + \epsilon_{it}, \quad i = 1, \dots, I, \quad t= 1, \dots, T \nonumber \\
	&\bm{\beta} \sim N( \bm \mu_{\beta} , \Sigma_{\beta}) \nonumber\\
	&\bm{w}_t|\bm{w}_{t-1} \sim N(\xi \bm{w}_{t-1}, \tau^2 Q(W,\rho)^{-1}), \quad t=2,...,T, \quad \bm{w}_1 \sim N(\mathbf{0}, \tau^2 Q (W,\rho)^{-1} ) 	\nonumber\\
	&\tau^2 \sim \text{Inv-Gamma}(3,2), \quad
	\sigma^2 \sim \text{Inv-Gamma}(3,2) \nonumber\\
	&\xi \sim \text{Uniform}(0,1), \quad  \rho \sim \text{Uniform}(0,1)\nonumber
\end{align}
The vector $\bm{\beta}$ is modeled as a multivariate Normal distribution with zero mean and the identity matrix as covariance matrix, the same prior specification used in the base distribution of the Dirichlet process in the proposed model. 

Using the data from Section~\ref{sec:data}, we compute and compare commonly used model comparison metrics, namely the out-of-sample root mean squared error (RMSE) and the out-of-sample mean absolute error (MAE). RMSE and MAE are computed using the $1$-year ahead forecast starting from the $5$-th year (2009), using all previous years as training set. The Bayesian models are additionally compared via the widely applicable information criterion \citep[WAIC,][]{watanabe2013widely} and the log marginal likelihood (LML). The marginal likelihood is obtained as the product of the one-step-ahead predictive likelihoods \citep[cf.][]{geweke2010comparing}:
\begin{align*}
 &p(\bm{Y}_{1:T}) = p(\bm{Y}_1) \prod_{t=2}^{T}  p(\bm{Y}_t|\bm{Y}_{1:t-1}) = \int_\Theta p(\bm{Y}_1|\bm\theta) p(\bm\theta) d\bm{\theta} \; \prod_{t=2}^{T} \int_{\Theta}  p(\bm{Y}_t| \bm{Y}_{1:t-1},\bm{\theta}) p(\bm{\theta} | \bm{Y}_{1:t-1}) d\bm{\theta}
\end{align*}
where $\bm{\theta}$ is the vector of all unknown parameters and $\Theta$ the corresponding parameter space. The logarithm of the marginal likelihood can therefore be decomposed into the sum of the logarithms of the one-step-ahead predictive likelihoods. To be consistent with the RMSE and MAE evaluation and to avoid excessive prior dependence, we report $p(\bm{Y}_{5:T}|\bm{Y}_{1:4})$, i.e., we begin the evaluation in the 5th year, treating data up to the 4th year as part of the prior information.

Following \cite{gelman2014}, the WAIC is obtained by calculating the log pointwise posterior predictive density (\ref{eq:waic_clppd}) and then adjusting for overfitting with the correction term (\ref{eq:waic_correct}): 
\begin{align}
&comp\_lppd= \sum_{i=1}^{I} \log \left(  \frac{1}{M} \sum_{m=1}^{M} p(\bm{Y}_i|\bm{\theta}^m) \right) \label{eq:waic_clppd}\\
&comp\_p_{WAIC}= 2 \sum_{i=1}^{I} \left(  \log \left(  \frac{1}{M} \sum_{m=1}^{M} p(\bm{Y}_i|\bm{\theta}^m) \right) - \frac{1}{M} \sum_{m=1}^{M} \log p(\bm{Y}_i|\bm{\theta}^m)  \right) \label{eq:waic_correct}\\
& WAIC =  -2 \cdot \left(comp\_lppd - comp\_p_{WAIC}\right) \nonumber
\end{align}
where $\bm{\theta}^m$ denotes the $m$-th posterior draw from a total of $M$ MCMC draws. Note that the metrics for the Bayesian models are computed using samples from the posterior. Therefore, they entail the posterior uncertainty for all parameters including, for the BSTC model, the estimated number of clusters.
The results summarized in Tables~\ref{tab:rmse} and \ref{tab:mae} show similar patterns for both RMSE and MAE. For the first half of the testing period, no model stands out as the clear winner. The IFE and BSTC models alternate in proving the least erroneous except for the year 2012, where the Pooled model, which does not capture any heterogeneity nor spatial correlation, performs best. Adding the SAR component mostly improves the estimates for the IFE and the Pooled model, and has almost no effect on the PS-ANOVA model. On average, the best frequentist models are the PS-ANOVA and the PS-ANOVA-SAR model, being outperformed only by the BSTC model. Table~\ref{tab:waic_lml} focuses on the Bayesian models, showing the logarithms of the one-step-ahead predictive likelihoods and the WAIC. Note that higher values of predictive likelihoods indicate better predictions, while higher values of WAIC indicate worse. The results resemble the RMSE and MAE, showing close metrics between the two models in the first years. As before, the BSTC model performs poorly in the prediction for the year 2012 and surpasses its competitor in the following years.

\begin{table}[t]
\caption[RMSE]{Out-of-sample RMSE of the competing models and the Bayesian spatio-temporal clustering model.}
	\label{tab:rmse}
	\centering
		\begin{tabular}{lcccccccccc}
			\hline
			\hline
			Model & 2009 & 2010 & 2011 & 2012 & 2013 & 2014 & 2015 & 2016 & 2017 & Average  \\ 
			\hline
Pooled & 1.055 & 0.949 & 0.841 &  \textbf{0.516} & 0.522 & 0.715 & 0.620 & 0.681 & 0.663 & 0.729 \\ 
IFE & \textbf{0.353} & 0.306 &  \textbf{0.268} & 0.585 & 0.680 & 0.741 & 0.520 & 0.471 & 0.532 & 0.495 \\ 
Pooled-SAR & 0.516 & 0.479 & 0.421 & 0.791 & 0.839 & 0.907 & 0.659 & 0.706 & 0.656 & 0.664 \\ 
IFE-SAR & 0.469 & 0.359 & 0.320 & 0.589 & 0.610 & 0.609 & 0.406 & 0.418 & 0.463 & 0.471 \\ 
PS-ANOVA & 0.364 & 0.316 & 0.293 & 0.608 & 0.510 &  \textbf{0.444} & 0.393 & 0.440 & 0.438 & 0.423 \\ 
PS-ANOVA-SAR & 0.364 & 0.316 & 0.293 & 0.608 & 0.510 &  \textbf{0.444} & 0.393 & 0.440 & 0.438 & 0.423 \\ 
ST.CARar & 0.417 & 0.377 & 0.439 & 0.825 & 1.005 & 1.151 & 0.350 & 0.366 & 0.775 & 0.634 \\ 
BSTC & 0.382 & \textbf {0.302} & 0.292 & 0.640 & \textbf{0.471} & 0.500 & \textbf{0.313} & \textbf{0.319} & \textbf{0.384} & \textbf{0.400} \\ 
\hline	
\end{tabular}
\end{table}

\begin{table}[t]	
\caption{Out-of-sample MAE of the competing models and the Bayesian spatio-temporal clustering model.}
	\label{tab:mae}
\centering
		\begin{tabular}{lcccccccccc}
			\hline
			\hline
		Model & 2009 & 2010 & 2011 & 2012 & 2013 & 2014 & 2015 & 2016 & 2017 & Average  \\ 
			\hline
Pooled & 0.957 & 0.865 & 0.775 & \textbf{0.428} & 0.386 & 0.524 & 0.450 & 0.477 & 0.482 & 0.594 \\ 
IFE & 0.254 & 0.238 & \textbf{0.196} & 0.474 & 0.554 & 0.578 & 0.400 & 0.373 & 0.421 & 0.387 \\ 
Pooled-SAR & 0.403 & 0.369 & 0.327 & 0.672 & 0.712 & 0.756 & 0.523 & 0.557 & 0.513 & 0.537 \\ 
IFE-SAR & 0.371 & 0.280 & 0.254 & 0.498 & 0.524 & 0.502 & 0.320 & 0.328 & 0.345 & 0.380 \\ 
PS-ANOVA & 0.287 & 0.241 & 0.216 & 0.463 & 0.379 & \textbf{0.317} & 0.306 & 0.334 & 0.329 & 0.319 \\ 
PS-ANOVA-SAR & 0.287 & 0.241 & 0.216 & 0.463 & 0.379 & \textbf{0.317} & 0.306 & 0.334 & 0.329 & 0.319 \\ 
ST.CARar & 0.295 & 0.268 & 0.323 & 0.617 & 0.730 & 0.832 & 0.264 & 0.267 & 0.596 & 0.466 \\ 
BSTC & \textbf{0.237} & \textbf{0.232} & 0.216 & 0.499 & \textbf{0.350} & 0.356 & \textbf{0.239} & \textbf{0.241} & \textbf{0.284} & \textbf{0.295} \\ 
\hline	
\end{tabular}
\end{table}

\begin{table}[t]	
\caption{Logarithms of the one-step ahead predictive likelihoods, their sum and the WAIC of the Bayesian models.}
	\label{tab:waic_lml}
\centering
	\begin{tabular}{lccccccccccc} 
\hline 
\hline 
Model & 2009 & 2010 & 2011 & 2012 & 2013 & 2014 & 2015 & 2016 & 2017 & Sum & WAIC\\ 
\hline 
ST.CARar & -63 & -54 & -43 & \textbf{-89} & -134 & -181 & -70 & -86 & -140 & -861 & 2013 \\ 
BSTC & \textbf{-56} & \textbf{-37} & \textbf{-42} & -172 & \textbf{-66} & \textbf{-125} & \textbf{-63} & \textbf{-69} & \textbf{-107} & \textbf{-737} & \textbf{-737} \\  
\hline 
\end{tabular} 

\end{table}


\section{Discussion and conclusions}  
\label{sec:conclusion}

We develop a novel Bayesian semiparametric model for spatio-temporal areal data and apply it to the evolution of Italian unemployment rates. The proposed approach aims at tackling the difficult challenge of identifying well-pooled spatial and temporal areal units to identify spatio-temporal patterns in the data. We do so by interweaving elements from spatial statistics and Bayesian nonparametrics.

The proposed model uses a Bayesian nonparametric prior, the DP, to cluster location-specific ($i$) autoregressive parameters driving the spatio-temporal random effects and ($ii$) regression parameters of the time varying predictors. This implies a Bayesian nonparametric model for clustering the areal-specific time series of the unemployment rates. Posterior inference is achieved via a tailored Markov chain Monte Carlo algorithm.
The fitted model is consequently used to study the evolution of unemployment rates of Italian provinces from 2005 to 2017. We compute estimates of their clustering structure by minimizing posterior expectations of Binder's loss and VI. In doing so, we obtain interpretable results and shed light on the unemployment structure in Italy. The proposed approach fares well in an extensive out-of-sample comparison against popular alternatives, further validating its applicability to the problem at hand.

The model could be further generalized in different directions.
First, it might be of interest to investigate prior distributions
that do not imply stationarity. For example, the non-stationarity in time could be modeled by means of a time-varying parameter (TVP) model. To avoid potential overfitting, recently developed shrinkage priors could be used \citep[e.g.][]{bitto,cadonna2020}. 
Second, 
one could employ covariate-informed partition models such as PPMx \citep{muller2011product}, where prior distributions additionally encourage the grouping of areas with similar covariate values. 
Third, the structure of the neighboring matrix $W$ could be treated as random and thus learned from the data.
To this end, it might be interesting to consider a prior for the precision matrix of the random effects based on a directed acyclic graph representation of the spatial dependence \citep{datta2019, codazzi2022gaussian}.

\bibliographystyle{cas-model2-names}
\bibliography{bib}

\clearpage
\begin{appendices}

\section{Spatio-temporal random effects}
\label{sec:st_randeff}
\subsection{Joint distribution}
\label{sec:joint_randeff}

Using the Markov property, the joint distribution of $\tilde{\bm w} = \left( {\bm w_1}, \dots,  {\bm w_T}\right)'$ can be expressed as
$p(\tilde{\bm w}) = p( {\bm w}_1) p( {\bm w}_2|  {\bm w}_1) \cdots p( {\bm w}_T|  {\bm w}_{T-1})$.
Since each $p({\bm w}_t| {\bm w}_{t-1})$ is Gaussian, and so is  $p( {\bm w}_1)$,  the joint distribution is a multivariate Gaussian distribution. For any multivariate Gaussian distribution with mean $ \tilde{ \bm \mu}$ and precision matrix $\Omega$, it holds that
\begin{align}
	\label{eq:to}
	-2 \log p(\tilde{\bm w}) =  \tilde{\bm w}' \Omega \tilde{\bm w} - 2 \tilde{\bm \mu}' \Omega \tilde{\bm w} + k, 
\end{align}
where $k$ is a term which does not contain $\tilde {\bm w}$. From (\ref{eq:ARCAR}) and the distribution on $ {\bm w}_1$, we have that
\begin{align}
	\label{eq:from}
	-2 \log p(\tilde{\bm w})=  \tau^{-2} \bm{ w}_{1}'  Q(\rho, W) \bm{ w}_1 + \tau^{-2} \sum_{t=2}^{N} (\bm{ w}_{t} - diag(\bm \xi) \bm{ w}_{t-1} )'Q(\rho, W) (\bm{ w}_t - diag(\bm \xi) \bm{ w}_{t-1}),
\end{align}
Matching the terms in  \eqref{eq:to}  and  \eqref{eq:from}, we obtain that $ \tilde{ \bm \mu} = \bm 0$ and that $\Omega$ is tri-block diagonal, with blocks of dimension $I\times I$. Specifically, 
\begin{align*} 
	&\Omega_{t,t}=  \tau^{-2}  Q(\rho, W) + \tau^{-2} diag(\bm \xi) Q(\rho, W) diag(\bm \xi), \\ 
	&\Omega_{T,T}= \tau^{-2} Q(\rho, W),\\
	&\Omega_{t, t-1}= -\tau^{-2} diag(\bm \xi) Q(\rho, W), \quad  t=2, \dots, T.
\end{align*}

\subsection{Full conditional distribution}
\label{sec:fullcond_randeff}
Due to the high dimension of the time-varying parameters, tailored algorithms are required. A way of dealing with datasets of such high dimensionality is to exploit the characteristics of sparse matrices, and in particular block diagonal and banded matrices. Important work in this direction has been done in \cite{Rue2001} and \cite{Mccausland2011}.
The joint full conditional of $\tilde{\bm w}$ given the data and the other parameters in the model, is a multivariate normal of dimension $I \times T$. The precision matrix is tri-block diagonal with blocks of dimension $I \times I$.
From Lemma 2.2. in \cite{Rue2005}, we get that
the full conditional is $N_{I\times T}\left({\Psi}^{-1} \bm c, {\Psi}^{-1}\right)$, where the precision matrix $\Psi$ is tri-block diagonal, with blocks of dimension $I\times I$. Specifically:
	\begin{align*}
		& \Psi_{t,t} = \frac{1}{\sigma^{2}} \mathbb I_{I\times I} + \tau^{-2} (Q(\rho, W)  + diag(\bm \xi^{\star}_{\bm s}) Q(\rho, W)diag(\bm \xi^{\star}_{\bm s})), \quad t = 1, \dots, T-1 \\
		&\Psi_{T, T} = \frac{1}{\sigma^{2}} \mathbb I_{I\times I} + \tau^{-2} Q(\rho, W)\\
		&\Psi_{t,t+1} = -  \tau^{-2} Q(\rho, W) diag(\bm \xi^\star_{\bm s}), \quad t = 1, \dots, T-1
	\end{align*}
and $\bm c=(\bm c_1,\dots, \bm c_T)'$, with each element $\bm c_t$ being
\begin{align*}
	&\bm c_t = \frac{1}{\sigma^{2}} \left(\bm Y_t - diag\left(\bm X_t \bm B_{\bm s}^\star \right)\right), \quad t = 1, \dots, T
\end{align*} 
Note that since all building blocks of $\Psi$ depend on $Q(\rho, W)$, all of them inherit the structure of the adjacency matrix $W$. In our case this means that all blocks $\Psi_{t,t}$ and $\Psi_{t, t+1}$, $t = 1, \dots, T-1$, are banded.

\subsection{Sampling}
\label{sec:sampling}
We use the algorithm described in Result 2.1 in \cite{Mccausland2011}. Unlike algorithms based on the Kalman filter, this method is based on the precision matrix and computationally more efficient.

First, we pre-compute the quantities $\Sigma_1, \dots, \Sigma_T$ and $m_1, \dots, m_T$, by iterating the following steps.
For $t=1, \dots, T$;
\begin{enumerate}
	\item Compute the Cholesky decomposition of $\Sigma_t^{-1}= \Lambda_t \Lambda_t^T$, where:
	\begin{itemize}
		\item for $t = 1$:\\
		$\Sigma_t^{-1} = \Psi_{1,1} = \frac{1} {\sigma^{2}} \mathbb I_{I \times I} + \tau^{-2} ( Q(\rho, W) +  (diag(\bm \xi^{\star}_{\bm s}) Q(\rho, W) diag(\bm \xi^{\star}_{\bm s}))$
		\item for $t = 2, \dots, T-1$:\\
		$\Sigma_t^{-1} = \Psi_{t,t} - \Psi_{t,t-1}' \Sigma_{t-1} \Psi_{t,t-1} = 	\\ 
		\frac{1} {\sigma^{2}} \mathbb I_{I \times I} + \tau^{-2} (Q(\rho, W)  + diag(\bm \xi^{\star}_{\bm s}) Q(\rho, W) diag(\bm \xi^{\star}_{\bm s})) - 
		\tau^{-4}diag(\bm \xi^{\star}_{\bm s})  Q(\rho, W)' \Sigma_{t-1} Q(\rho, W) diag(\bm \xi^{\star}_{\bm s})$
	\end{itemize}
	Note that each matrix is banded, so we can use the appropriate Cholesky decomposition
	
	\item Compute $\Lambda_t^{-1}=  \left(-\tau^{-2}diag(\bm \xi^\star_{\bm s})Q(\rho, W)\right)$ via back-substitution. Again, we can use the algorithm for banded  matrices
 
	\item Compute  $\tau^{-4} diag(\bm \xi^\star_{\bm s})  Q(\rho, W)'  \Sigma_{t-1}  Q(\rho, W) diag(\bm \xi^\star_{\bm s}) = \\
	\tau^{-4} diag(\bm \xi^\star_{\bm s})(\Lambda_{t-1}^{-1} Q(\rho, W))'\Lambda_{t-1}^{-1} Q(\rho, W)diag(\bm \xi^\star_{\bm s})$ 
 
	\item Compute $m_t$ using triangular back-substitution twice, $m_1= (\Lambda_1 ')^{-1}(\Lambda_1^{-1} c_1)$ and $m_t  = (\Lambda_t')^{-1}(\Lambda_t^{-1}(c_t + \tau^{-2}  diag(\bm \xi^\star_{\bm s})  Q(\rho, W)'m_{t-1}))$,
	for $t=2, \dots, T$
\end{enumerate}
To draw $\bm w \sim (\Psi^{-1} \bm c, \Psi^{-1})$, we proceed backwards. For $ t=T, \dots, 1$:
\begin{enumerate}
	\item Sample $\bm z_t \sim N(\bm 0, \mathbb I_{I \times I})$, which is equivalent to sampling $I$ independent standard normals
	\item Compute $\bm w_t$ using matrix multiplication and back-substitution, where $\bm w_T = m_T + (\Lambda_T)^{-1} \bm z_T$, and  $\bm w_t = m_t + (\Lambda_t)^{-1} (\bm z_t + \Lambda_t^{-1} \tau^{-2} diag(\bm \xi^\star_{\bm s}) \bm w_{t+1})$
\end{enumerate}

\section{Full conditional distributions for the BNP part and the remaining parameters}
\label{sec:fullcond_bnp_remain}
\subsection*{Full conditionals for $\bm{s}$}
We start by removing the $i$-th allocation variable $s_i$ from the vector $\bm s$ and by denoting the remaining vector as $\bm s^{-i} = \left(s_1, \dots, s_{i-1}, s_{i+1}, \dots, s_I\right)$. Similarly, we can re-write the $j$-th cluster, after removing $s_i$, as $C_j^{-i}$ and the associated cluster size as $n_j^{-i} = |C_j^{-i}|$. The number of clusters is then denoted as $K^{-i}_I$. Following from the expression of the P\'{o}lya Urn prior and the assumption of exchangeability among the element of $\bm s$, the probability of allocating the removed observation to an existing or to a new cluster are, respectively:
\begin{align}\label{eq:Allocation_DDP}
	&p\left(s_i = j | \bm s^{-i}, \bm y, \bm x, \bm \beta^\star, \bm \xi^\star, \bm w, \sigma^2 \right) \propto \\
	&\left\{
	\begin{array}{ll}
		n_j^{-i} p\left(\bm y_i | \bm x_i, \bm \beta^\star_j, \bm w_i, \sigma^2 \right) p\left(\bm w_i | \bm w^{-i}, \xi^\star_j, \bm \xi^\star_{\bm s^{-i}}, \tau^2, \rho, W\right), & j = 1, \dots, K_I^{-i}\\
		\alpha \int p\left(\bm y_i | \bm x_i, \bm \beta^\star, \bm w_i , \sigma^2\right) p\left(\bm w_i | \bm w^{-i}, \xi^\star, \bm \xi^\star_{\bm s^{-i}}, \tau^2, \rho, W\right) dP_0\left(\bm \beta^\star, \xi^\star\right), & j = K_I^{-i} + 1 \nonumber
	\end{array}
	\right.
\end{align}
where we indicate with $\bm w^{-i}$ the matrix of variables $\bm w$ after removing the $i$-th row. Before giving the expression of the above allocation probabilities, we point out that the integral involved in this calculation is available in closed form only in case of conjugacy. To avoid computing the integral above, we resort to the algorithm of \cite{FavTeh13}, implementing an extension of the popular Algorithm 8 of \cite{Neal00} which includes a re-use step. The second line of \eqref{eq:Allocation_DDP} becomes:
\begin{equation*}
	\frac{\alpha}{N^{aux}} p\left(\bm y_i | \bm x_i, \bm \beta^{aux}_l, \bm w_i\right) p\left(\bm w_i | \bm w^{-i}, \xi^{aux}_l, \bm \xi^\star_{\bm s^{-i}}, \tau^2, \rho, W\right), \quad l = 1, \dots, N^{aux}
\end{equation*}
where $\left(\bm \beta, \bm \xi\right)^{aux} = \left(\bm \beta^{aux}_l, \xi^{aux}_l\right)_{l = 1}^{N^{aux}}$ is a vector of $N^{aux}$ auxiliary variables i.i.d. from $P_0$. The number of auxiliary variables $N^{aux}$ to be used in the algorithm is set to $N^{aux} = 20$.

The allocation probabilities in \eqref{eq:Allocation_DDP} contain the product of the likelihood of the data $\bm y_i$ and the spatio-temporal random effects $\bm w_i$ for areal unit $i$. The former is the sampling model proposed in \eqref{eq:full_model}, for which conditional independence applies, yielding:
\begin{equation*}
	p\left(\bm y_i | \bm x_i, \bm \beta^\star_j, \sigma^{2}, \bm w_i\right) = N_T\left(\bm y_i | \bm x_i \bm \beta^\star_j + \bm w_i, \sigma^2\mathbb I_T \right)
\end{equation*}
To compute the latter we use the VAR(1) structure of $\bm w$:
\begin{equation*}
p\left(\bm w | \bm \xi^\star_j, \tau^2, \rho, W\right) = N\left(\bm w_1 | \bm 0, \tau^2Q^{-1}\left(\rho, W\right)\right) \prod_{t = 2}^T N\left(\bm w_t | diag\left(\bm \xi^\star_{\bm s}\right) \bm w_{t-1}, \tau^2Q^{-1}\left(\rho, W\right)\right)
\end{equation*}
For each time point $t = 1, \dots, T$, we consider the conditional distribution of the $i$-th component of the vector $\bm w_t$. Denote by $w_{it}$ such component, and by $\bm w^{-i}_{t}$ the rest of the vector $\bm w_t$, and let $\setminus i = \{1, \dots, i-1, i+1, \dots, I\}$. Further conditioning on $\bm w_{t-1}$ yields:
\begin{align*}
	&p\left( w_{it}|  \bm w^{-i}_{t}, \bm w_{t-1}, \tau^2, \rho, \bm s^{-i}, \xi^\star_j, \bm \xi^\star_{\bm s^{-i}}\right) = N\left(w_{it} | \mu^c_{it}, {\sigma_i^2}^{c}\right)\\
	&\text{where}\\
	& \mu^c_{it} = \xi^\star_j w_{i t-1} -{\sigma_i^2}^{c} Q_{i \setminus i} / \tau^2 \left({{\bm w_{t}^{-i}}} - diag\left(\bm \xi^\star_{\bm s^{-i}}\right) \bm w_{t-1}^{-i}\right)\\
	&{\sigma_i^2}^{c} = \tau^2 /Q_{ii}
\end{align*}
Hence
\begin{equation*}
	p(\bm w_i | \bm w^{-i}, \xi^\star_j, \bm \xi^\star_{\bm s^{-i}}, \tau^2, \rho, W) \propto N\left(w_{i1} | \mu^c_{i1}, {\sigma_i^2}^{c}\right) \prod_{t = 2}^T N\left(w_{it} | \mu^c_{it}, {\sigma_i^2}^{c}\right)
\end{equation*}

\subsection*{Full conditionals for $\bm \beta^\star,  \bm \xi^\star$, and $\alpha$}
Notice that, when a new cluster is created, one of the auxiliary variables $(\bm \beta, \bm \xi)^{aux}$ is selected as new component of the vector of unique values. However, the whole set of unique values $(\bm \beta^\star, \bm \xi^\star)$ can be updated via Gibbs or Metropolis-Hastings steps. 

We can exploit the conjugacy of the full conditionals of $\bm \beta^\star_j$, for $j = 1, \dots, K_i$, obtaining:
\begin{align*}
	& p(\bm \beta^\star_j |  \cdot) = N_{p+1}(\bm \beta^\star_j | \overline{m}_j, \overline{S}_j) \\
	& \overline{S}_j = \left( \Sigma^{-1}_{\bm \beta} + \sum_{i \in C_j} \bm x_i^{\top} \bm x_i / \sigma^{2} \right)^{-1} \\
	& \overline{m}_j = \overline{S}_j \left( \Sigma^{-1}_{\bm\beta} \bm \mu_{\bm \beta} + \sum_{i \in C_j} \bm x_i^{\top} \left(\bm y_i - \bm w_i\right)/\sigma^{2} \right)
\end{align*}

The update of $\xi^\star_j$ is performed using a Metropolis-Hastings step targeting the following full-conditional, up to a normalizing constant:
\begin{align*}
	& p\left(\xi^{\star}_j | \cdot\right) \propto \text{Beta}_{(-1,1)}\left(\xi_j^{\star} | a_{\xi}, b_{\xi}\right) \prod_{t=2}^T N\left({\bm w}_{C_j t} | \mu^c_{j t}, \Sigma^c_{j t}\right) \\
	& \mu^c_{j t} = \xi^{\star}_j{\bm w}_{C_j t-1} - \Sigma^c_{j t} Q_{C_j \setminus C_j} / \tau^2 \left({\bm w}_{C_j t} - diag\left(\bm \xi^{\star}_{s \neq j}\right){\bm w}_{\setminus C_j t-1}\right) \\
	&\Sigma^c_{j t} = \tau^2 Q_{C_j C_j}^{-1}
\end{align*}
where ${\bm w}_{C_j t}$ denotes the spatio-temporal random effects of the provinces belonging to the $j$-th cluster at time point $t$. 

Using a gamma prior for $\alpha$ and introducing an auxiliary variable $x$ such that
$ x|\alpha,K \sim Beta(\alpha +1 , n)  $, the full conditional of $\alpha $ reduces to a mixture of two gamma densities \citep[][]{west1992hyperparameter}:
\begin{align*}
   \alpha | x,K \sim \pi_x Gamma(\alpha+K, b - \log(x))+ (1-\pi_x) Gamma(\alpha+K-1,b-\log(x))
\end{align*}
with the weights $\pi_x$ defined through
\begin{align*}
   \frac{\pi_x}{1-\pi_x}=\frac{\alpha+1}{n(b-log(x))}
\end{align*}

\subsection*{Full conditionals for $\sigma^2, \tau^2$, and $\rho$}
Staying in a conjugate framework the full conditional distributions for $\sigma^2$ and $\tau^2$ are both inverse-gamma distributions. The parameters for $\sigma^2$ are
\begin{align*}
	&\overline{a}_{\sigma^2} = a_{\sigma^2} + IT/2\\
	&\overline{b}_{\sigma^2} = b_{\sigma^2} + \dfrac{1}{2} \sum_{i=1}^I\sum_{t=1}^T (y_{it} - (\bm x_{it}' \bm \beta^\star_{s_i} + w_{it} ))^2
\end{align*}
and for $\tau^2$
\begin{align*}
	&\overline{a}_{\tau^2} = a_{\tau^2} + IT/2\\
	&\overline{b}_{\tau^2} = b_{\tau^2} + \dfrac{1}{2} 
	\left (
	\bm w'_1 Q(\rho, W) \bm w_1 + \sum_{t=2}^T (\bm w_t- diag(\bm \xi^\star_{\bm s}) \bm w_{t-1})'Q(\rho, W) (\bm w_t- diag(\bm \xi^\star_{\bm s})\bm w_{t-1}) \right )
\end{align*}
The full conditional distribution of the  spatial autoregressive parameter $\rho$ takes the form
\begin{align*}
		&\text{Beta} \left(\rho | \alpha_{\rho}, \beta_{\rho}\right) \times
		\mid Q(\rho, W) \mid^{T/2} \exp \left\{-\dfrac{1}{2\tau^2} ( \bm w'_1 Q(\rho, W) \bm w_1 \right\} \times \\
		&\prod_{t=2}^T \exp \left\{-\dfrac{1}{2\tau^2} (\bm w_t -diag(\bm \xi^\star_{\bm s}) \bm w_{t-1})' Q(\rho, W) (\bm w_t - diag(\bm \xi^\star_{\bm s}) \bm w_{t-1})
		\right\}
\end{align*}
and is sampled using a Metropolis-Hastings step with a  random walk on the logit transformation of $\rho$. 

\section{Sensitivity analysis of different methods of loss calculation}
\label{sec:sens_loss}
In this section, we perform a sensitivity analysis to assess the effect of the choice of loss function in estimating the partition of the areal units in the proposed model. Specifically, we consider the well-known Binder and Variation of Information loss functions.

The Binder loss function takes the form
\begin{align*}
	L_{Binder}(\bm{s},\bm{\hat{s}})=    	\sum_{i<j}\left(a \cdot \mathbb{1}_{\{s_i=s_j\}}\mathbb{1}_{\{\hat{s}_i\neq\hat{s}_j\}}
	+b \cdot \mathbb{1}_{\{s_i\neq s_j\}} \mathbb{1}_{\{\hat{s}_i=\hat{s}_j\}}\right)	
\end{align*}
where $\bm{s}=(s_1,...,s_n)$ is a vector of true cluster labels and $\bm{\hat{s}}$ its estimate. The true and estimated partitions are denoted by $\mathcal{C}$ and $\hat{\mathcal{C}}$, respectively. Note that the cost incurred by incorrectly separating two items is denoted by $a$ and the cost of incorrectly clustering two items together by $b$. It is important to note that the actual choice of $a$ and $b$ can have a substantial impact on the number of clusters and has to be chosen with care \citep[see also][]{salso-paper}. \cite{lau2007} showed that the partition that minimizes the posterior expectation of the Binder loss equals the one that maximizes 
\begin{align*}
f(\bm{\hat{s}})=\sum_{i<j}\mathbb{1}_{\{\hat{s}_i=\hat{s}_j\}} \left(S_{i,j}-\frac{b}{a+b} \right)
\end{align*}
where the so-called posterior similarity matrix $S_{i,j}$ denotes the posterior probability of areal unit $i$ and areal unit $j$ belonging to the same cluster.

\cite{salso-paper} propose to minimize the posterior expectation of the Binder loss using a stochastic search algorithm, implemented in the R package \texttt{salso} \citep{salso-package}.
We refer to the two methods as $L_{Binder\_LG}$ and $L_{Binder\_salso}$, respectively.

The generalized variation of information \citep[GVI,][]{salso-paper} takes the form
\begin{align}
	L_{GVI}(\mathcal{C},\hat{\mathcal{C}})=  
	a \sum_{C \in \mathcal{C}} \frac{|C|}{n} \text{log}_2\left( \frac{|C|}{n}  \right)+	
	b \sum_{\hat{C} \in \hat{\mathcal{C}}} \frac{|\hat{C}|}{n} \text{log}_2\left( \frac{|\hat{C}|}{n}  \right)-
	(a+b) \sum_{C \in \mathcal{C}}\sum_{\hat{C} \in \hat{\mathcal{C}}} 	\frac{|C \cap \hat{C}|}{n} \text{log}_2
		\left( \frac{|C \cap \hat{C}|}{n}  \right)
		\label{gvi}
\end{align}
where ${|C|}$ and ${|\hat{C}|}$ are the cluster sizes of elements belonging to the true and estimated partitions, respectively. The first term in Eq.~(\ref{gvi}) represents the negative individual entropy of the true partitition, scaled by the misclassification cost $a$. The second term represents the negative individual entropy of the estimated partitition, scaled by the misclassification cost $b$. The last term represents their joint entropy scaled by the mean of $a$ and $b$, $(a+b)/2$. The range of the individual entropy goes from $0$ to $log_2(n) \approx 6.78$, while the joint entropy ranges from $0$ to $2log_2(n) \approx 13.56$.
We use the entropy of the sampled partitions to quantify the variability in the posterior distribution of the partition. We observe low values of the individual entropy, with a posterior mean of 2.1 and a $95\%$ credible interval of (1.6, 2.6), indicating low variability in the partitions explored by the MCMC chain. We quantify the average deviation of the chosen partition and the sampled partitions using their joint entropy, resulting in a posterior mean of 5.1 and a $95\%$ credible interval of (3.4, 7.0), suggesting that the partitions explored by the MCMC algorithm are similar to the one presented in the application.

To assess how well the model retains predictive accuracy under the different partitions, we compute the WAIC, in-sample RMSE and in-sample MAE. The RMSE and MAE are calculated by running the model on the whole dataset and comparing the predicted response variable $\hat{y}_{i,t} = \bm{x}_{i,t}\hat{\bm \beta}^\star_{s_i}+\hat{w}_{i,t}$, $t = 1, \dots, T$, $i = 1, \dots, I$, with the corresponding true value. These goodness-of-fit metrics are calculated using posterior averages of $\hat{\bm \beta}^\star_{s_i}$ and $\hat{\bm \xi}^\star_{s_i}$, conditionally on the partitions estimated using different loss function specifications (Binder and GVI loss, with varying $a$ and $b = 1$). All estimated partitions are computed using posterior draws of the BSTC method described in Section \ref{subsec:compmod}. The results are summarized in Table~\ref{tab:lf_comparison}, where lower values of WAIC, RMSE, and MAE imply higher predictive accuracy. The partitions obtained with $L_{Binder\_LG}$ are the most consistent with regard to the estimated number of clusters $\hat{K}_I$, yielding 7 or 6 clusters in the majority of cases. The other two methods show more sensitivity to the choice of $a$, with $L_{GVI}$ rapidly decreasing the number of clusters for higher values of $a$.
\begin{table}[t!]
\centering 
\caption{Estimated number of clusters $\hat{K}_I$, goodness-of-fit metrics (WAIC, RMSE, and MAE) and Rand Index for different combinations of loss function and misclassification cost $a$. The latter compares the different estimated partitions with the one presented in the manuscript ($L_{Binder\_LG}$, $a = 1$, marked in bold).}
\label{tab:lf_comparison}
\setlength{\tabcolsep}{15pt} 
\begin{tabular}{ c  c c c c c c} 
\hline
Method of loss calculation & Cost $a$ & $\hat{K}_I$ & WAIC & RMSE & MAE & Rand Index\\ 
\hline
\multirow{5}{0.15\columnwidth}{ $L_{Binder\_LG}$} 
& 0.5 &  8 &  6101 & 0.273 & 0.199 & 0.90 \\ 
&  \textbf{1} &   \textbf{7} &   \textbf{6726} &  \textbf{0.246} &  \textbf{0.185} &  \textbf{1.00} \\ 
& 1.5 &  7 &  6879 & 0.254 & 0.192 & 0.91 \\ 
& 2 &  6 & 10478 & 0.383 & 0.258 & 0.83 \\ 
& 3 &  6 &  9966 & 0.379 & 0.258 & 0.82 \\ 
\hline
\multirow{5}{0.15\columnwidth}{$L_{Binder\_salso}$ }  
& 0.5 & 11 &  8808 & 0.259 & 0.186 & 0.90 \\ 
& 1 & 11 &  7658 & 0.246 & 0.180 & 0.95 \\ 
& 1.5 & 11 &  4245 & 0.230 & 0.181 & 0.88 \\ 
& 2 &  9 &  6215 & 0.282 & 0.218 & 0.82 \\ 
& 3 &  6 &  7927 & 0.308 & 0.228 & 0.81 \\
\hline
\multirow{5}{0.15\columnwidth}{ $L_{GVI}$ } 
& 0.5 & 11 &  5244 & 0.251 & 0.202 & 0.83 \\ 
& 1 &  8 &  6971 & 0.312 & 0.256 & 0.70 \\ 
& 1.5 &  5 &  9545 & 0.354 & 0.284 & 0.64 \\ 
& 2 &  3 & 11832 & 0.405 & 0.317 & 0.60 \\ 
& 3 &  3 & 11832 & 0.405 & 0.317 & 0.60 \\ 
\hline
\end{tabular} 
\end{table}
%
%
%
%
Note that in most cases, the comparison metrics favor partitions with a higher number of clusters. However, too many different clusters exacerbate meaningful economic interpretation. For this reason, the choice of a representative partition requires a trade-off between a number of clusters that is high enough to capture individual heterogeneity and sparse enough to be economically interpretable. Following this reasoning, we choose the partition obtained using the $L_{Binder\_LG}$ and $a = 1$. The final column Table~\ref{tab:lf_comparison} reports the Rand Index, a similarity measure for partitions, where values closer to 1 indicate partitions that are more similar. The values indicate high consistency of the different partitions, especially ones obtained using the Binder loss function.

\section{Alternative model specifications}
\label{sec:alt_model_spec}
 
This section shows the results of a performance comparison of the model proposed in Section~\ref{sec:Model} with alternative specifications.
In particular, (a) considers a finite-dimensional version of the nonparametric prior, while (b) focuses on different options to include the DP as a prior for the parameters $\bm \beta$ and $\xi$. All other prior specifications not concerning these two modifications are kept as in Section~\ref{sec:prior_elicit}.

In (a), we use a finite mixture model, instead of an infinite one.
\cite{mfm} point out that both Dirichlet process mixtures (DPMs) and mixtures of finite mixtures (MFMs) share several properties essential in the implementation of state-of-the art sampling algorithms and use this to introduce an alternative version of Neal's popular Algorithm 8, suited for MFMs. By applying their method, we estimate a finite analogue of the proposed model (BSTC-MFM). We follow \cite{mfm} and choose a geometric prior with success probability equal to 0.1 for the number of components.

Turning to (b), we examine a model where we employ one DP prior to cluster the $\bm \beta$ coefficients and a separate one for the temporal autoregressive parameter $\xi$. In other words, we examine a model where $\xi$ and $\bm \beta$ are clustered using independent DP priors. This approach is termed BSTC-2DPs. In addition, we examine a model where $\xi$ is removed from the clustering process altogether, and instead modelled via a parametric prior. This approach is termed BSTC-XI. In both cases, the \text{Beta}$_{(-1,1)}(a,b)$ prior is used, in the first case as the base measure of the now separate DP prior and in the second one as the parametric prior.

While we acknowledge that a rigorous comparison of different prior specifications is beyond the scope of the current work, these experiments can nevertheless shed light on the model's sensitivity to prior modifications. To this end, we compare the different model specifications by analyzing different partitions obtained from samples of each respective model. The MCMC sampler is run for the same number of iterations and burn-in as specified in Section~\ref{sec:econint}.

\begin{table}[t!] 
\centering 
  \caption{Summary results of the estimated partitions under different model and loss function specifications. } 
  \label{tab:alt_model_clustcomp} 
\begin{tabular}{@{\extracolsep{5pt}} lccccc} 
\hline \\[-1.8ex] 
\vtop{\hbox{\strut Model}\hbox{\strut }}  & \vtop{\hbox{\strut Method of}\hbox{\strut loss calculation}}   & $\hat{K}_I$ & \vtop{\hbox{\strut Size of North- }\hbox{\strut Central Cluster}}   &\vtop{\hbox{\strut Size of}\hbox{\strut Sicilian Cluster}}    &   \vtop{\hbox{\strut Medio-Campidano}\hbox{\strut as singleton}}\\ 
\hline \\[-1.8ex] 
BSTC & \multirow{4}{0.15\columnwidth}{ $L_{Binder\_LG}$}  &  7 & 55 & 9 & TRUE \\ 
BSTC-MFM & &  8 & 50 & 9 & FALSE \\ 
BSTC-2DPs &  &  9 & 59 & 8 & TRUE \\ 
BSTC-XI &  &  8 & 56 & 7 & TRUE \\ 
\hline
BSTC & \multirow{4}{0.15\columnwidth}{ $L_{Binder\_salso}$}  & 11 & 54 & 8 & FALSE \\ 
BSTC-MFM &  & 12 & 49 & 8 & FALSE \\ 
BSTC-2DPs & & 11 & 56 & 8 & FALSE \\ 
BSTC-XI &  & 11 & 53 & 8 & FALSE \\ 
\hline
BSTC & \multirow{4}{0.15\columnwidth}{ $L_{GVI}$}   &  8 & 83 & 8 & TRUE \\ 
BSTC-MFM & &  9 & 79 & 8 & TRUE \\ 
BSTC-2DPs & &  8 & 82 & 9 & TRUE \\ 
BSTC-XI & &  7 & 84 & 9 & TRUE \\ 
\hline \\[-1.8ex] 
\end{tabular} 
\end{table}

\begin{table}[t!] \centering 
  \caption{WAIC, in-sample RMSE, and in-sample MAE for alternative model specifications.} 
  \label{tab:alt_mod_metrics} 
\begin{tabular}{@{\extracolsep{5pt}} lccc} 
\hline \\[-1.8ex] 
Model & WAIC & RMSE & MAE \\ 
\hline \\[-1.8ex] 
BSTC & -736.528 & 0.143 & 0.114 \\ 
BSTC-MFM & -772.119 & 0.139 & 0.111 \\ 
BSTC-2DPs & -738.459 & 0.145 & 0.117 \\ 
BSTC-XI & -745.440 & 0.144 & 0.117 \\ 
\hline \\[-1.8ex] 
\end{tabular} 
\end{table} 

Table \ref{tab:alt_model_clustcomp} shows some essential characteristics of the partitions obtained by applying the loss-based methods introduced in Section~\ref{sec:sens_loss} with $a=1$ to MCMC samples from the alternative models. We compare the estimated number of clusters $\hat{K}_I$, the size of the biggest cluster, located in the North and Center, the size of the cluster predominantly covering Sicily, and whether the province Medio-Campidano was classified as a singleton cluster. Since a decisive part of the parameter estimation depends on the visited partitions, these characteristics together with the comparison metrics reported in Table~\ref{tab:alt_mod_metrics} serve as a proxy for measuring the similarities between the alternative models. Additionally, we compute the WAIC, in-sample RMSE and in-sample MAE on the complete dataset, again using the sampled posterior values and no summarisation method.

The estimated number of cluster $\hat{K}_I$, the size of the Sicilian cluster as well as the inspected singleton cluster hardly differ among the different models. Regarding the size of the biggest cluster, we see some variation between model specifications, but varying strongly between the chosen method of loss calculation. The results show strong consistency in estimating the partition of the provinces among the alternative model specifications. Table \ref{tab:alt_mod_metrics} underlines this consideration, showing almost no distinction in goodness-of-fit metrics between the models.
Each of the examined alternative model specifications provides a different and valuable view on the underlying process. Nevertheless, the analysis shows that their impact on the main aspect of this work, i.e. the model based clustering of the Italian provinces, is small.

\section{Simulation study}
\label{sec:sim_study} 
\begin{figure}[t!]
	\centering
		\includegraphics[width=.6\textwidth]{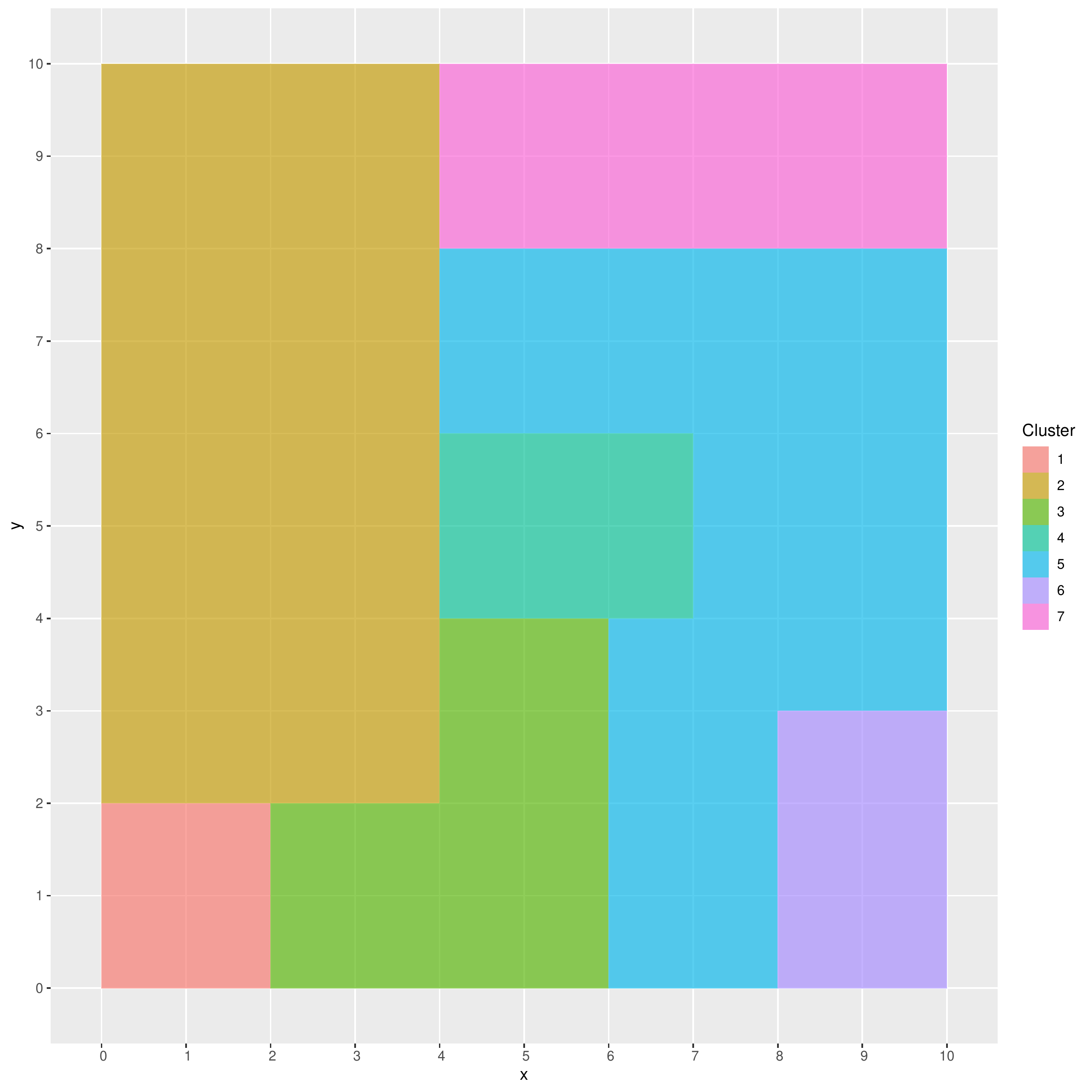}
		\caption{Underlying cluster structure for simulated data.}
		\label{fig:scen2_map}
\end{figure}

We present a simulation study to assess the ability of the proposed model to correctly estimate the parameters and the clustering of the areal units. Using the model introduced in Section~\ref{sec:Model}, we simulate data for a grid of $10 \times 10$ areal units. We define seven clusters of well separated groups of adjacent units, as depicted in Figure~\ref{fig:scen2_map}.
Each cluster is assigned three covariates and four regression coefficients, all of which are independently sampled from a standard normal distribution, and each cluster's $\xi$ is sampled from a standard uniform distribution. The spatial autocorrelation parameter $\rho$ is set to 0.95, while $\sigma^2 = \tau^2 = 1$. The prior settings are the same as described in Section~\ref{sec:prior_elicit}, except for a uniform prior for $\rho$.

\begin{figure}[t!]
	\centering
		\includegraphics[width=1\textwidth]{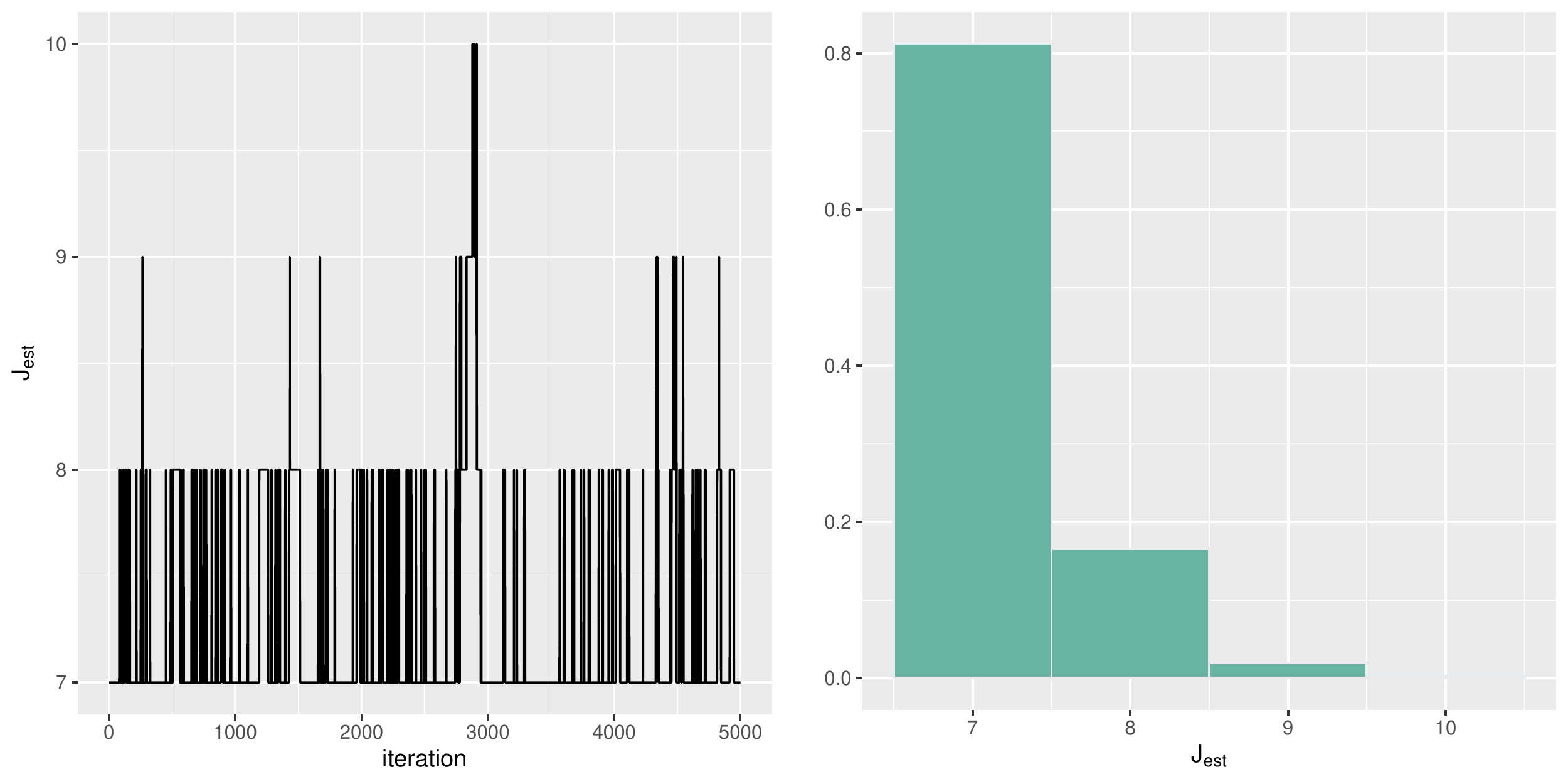}
		\caption{Traceplot (left-hand side) and posterior distribution (right-hand side) of $K$.}
		\label{fig:scen2_J_plots}
\end{figure}

\begin{figure}[!t]
	\centering
		\includegraphics[width=1.0\textwidth]{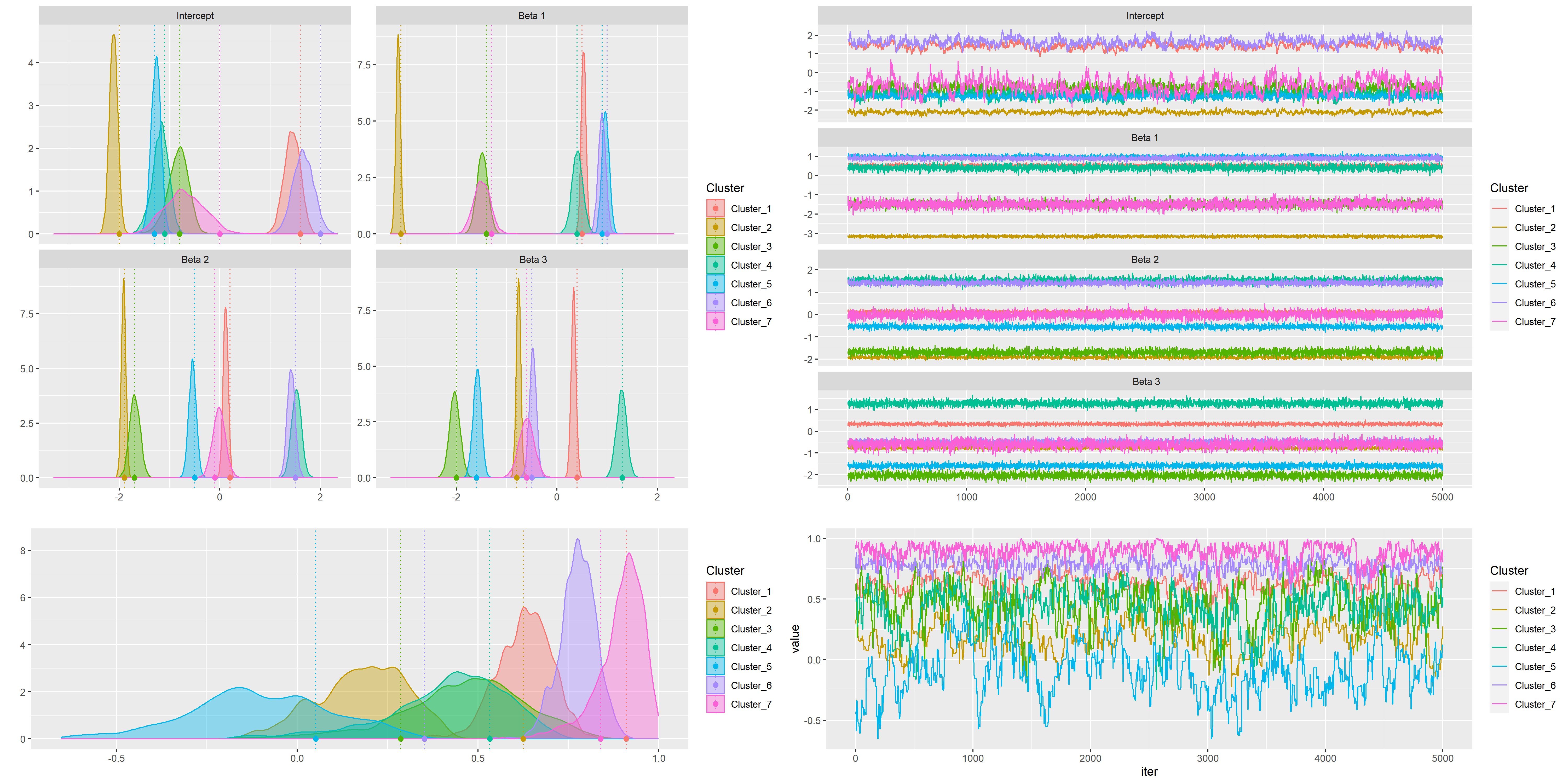}
		\caption{Density and traceplots of the cluster-specific regression coefficients $\bm \beta$ and the autoregressive coefficients $\bm \xi$.}
		\label{fig:scen2_beta_xi_plots}
\end{figure}

 We run the MCMC sampler for 10,000 iterations and discard the first 5,000 as burn-in. Figures~\ref{fig:scen2_J_plots} and \ref{fig:scen2_beta_xi_plots} show posterior inference for one of these datasets. We observe in Figure~\ref{fig:scen2_J_plots} that the sampler visits up to 12 clusters but coincides with the true value almost $80\%$ of the times. The density plots show that the true $\bm \beta$ coefficients are accurately recovered, while the trace plots imply good mixing and convergence.

\begin{figure}[t!]
	\centering
		\includegraphics[width=1.0\textwidth]{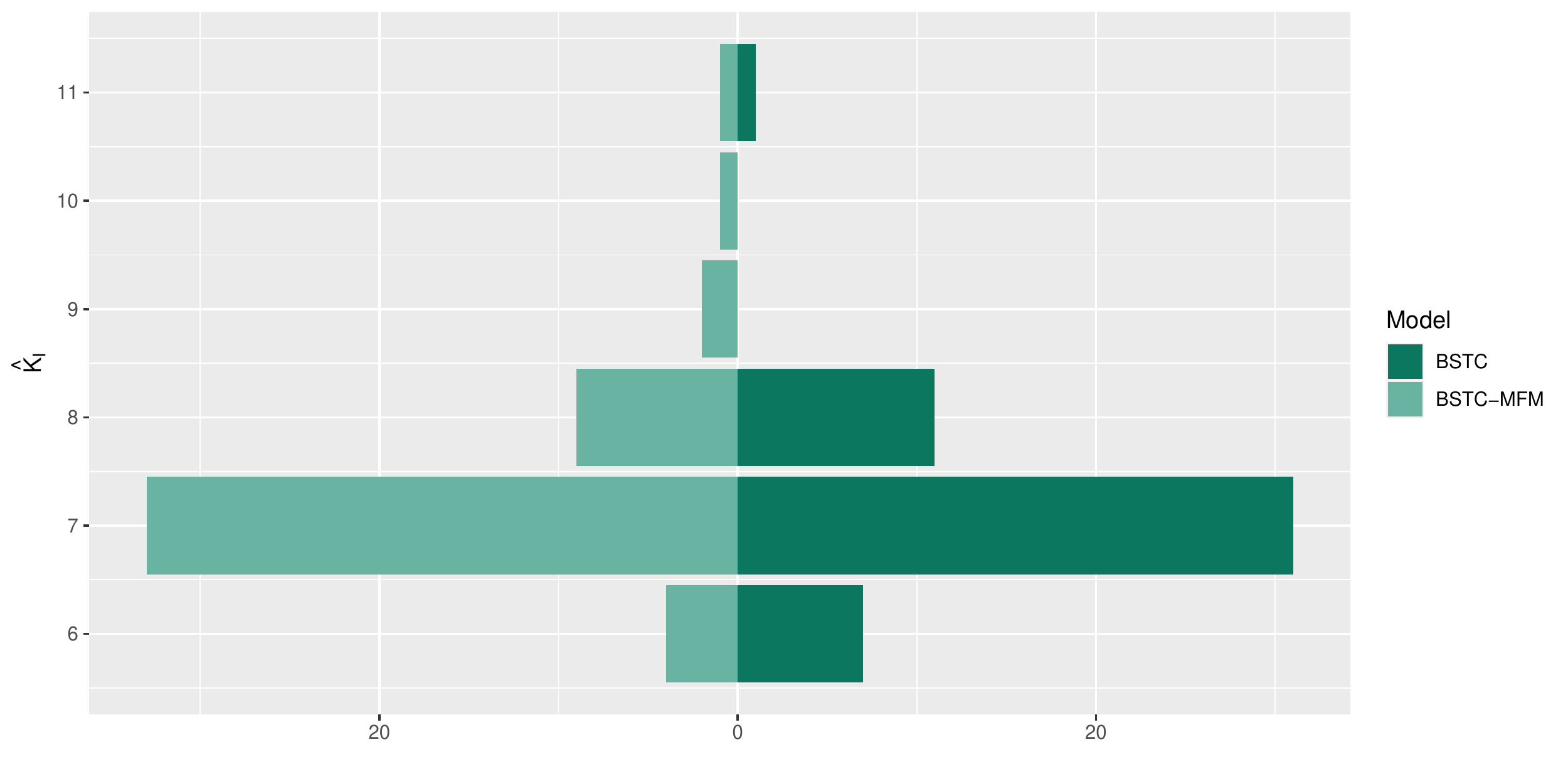}
		\caption{Barplots for $\hat{K}_I$ obtained from the 50 simulated datasets for the BSTC-MFM (left) and BSTC (right) models. The true value for the simulated data is $K=7$.}
		\label{fig:ss_barplot}
\end{figure}

To guarantee robustness of the results, and in addition to compare the DP process prior with the MFM alternative, we simulate 50 replicated datasets using different random seeds and estimate both the BSTC as well as the BSTC-MFM specification on these data. Figure~\ref{fig:ss_barplot} displays the estimated number of clusters over the 50 replicate datasets, obtained using the Binder loss function with misclassification cost $a = 1$, for both models. We can observe that the MFM model recovers the true number of clusters slightly more often than the DPM model, but also deviates further from it, sometimes estimating nine or ten clusters. Although both models find 11 clusters in one case, they do so for the same dataset which seems to prove a particularly difficult instance of the data generating process. It is important to emphasize that out of 50 different datasets, the two models agree on the number of clusters in 38 of them and only deviate by one cluster in the majority of the remaining cases.
Overall, both models show very similar cluster estimates, even though the prior distributions for the number of clusters are not identical \citep[cf.][]{heretoinfinity}.

\end{appendices}

\end{document}